\documentclass[%
 aip, amsmath,amssymb,reprint]{revtex4-1}
\pdfoutput=1 

\newcommand{\nnm}{\nonumber} 

\newcommand{\ra}{\rangle}
\newcommand{\la}{\langle}
\newcommand{\upa}{\uparrow}
\newcommand{\dna}{\downarrow}
\usepackage{graphicx}

\usepackage{epsfig}

\begin{document}

\title{Effects of valley splitting on resonant-tunneling readout of spin qubits}%

\author{Tetsufumi Tanamoto}
\affiliation{Department of Information and Electronic Engineering, Teikyo University,
Toyosatodai, Utsunomiya  320-8511, Japan} 
\email{tanamoto@ics.teikyo-u.ac.jp}


\author{Keiji Ono}
\affiliation{Advanced device laboratory, RIKEN, Wako-shi, Saitama 351-0198, Japan}

\begin{abstract}
The effect of valley splitting on the readout of qubit states is theoretically investigated in a three-quantum-dot (QD) system. 
A single unit of the three-QD system consists of qubit-QDs and a channel-QD that is connected to a conventional transistor. 
The nonlinear source--drain current characteristics under resonant-tunneling effects are used to distinguish different qubit states. 
Using nonequilibrium Green functions, 
the current formula for the three-QD system is derived when each QD has two valley energy levels. 
Two valley states in each QD are considered to be affected by variations in the fabrication process. 
We found that when valley splitting is smaller than Zeeman splitting, 
the current nonlinearity can improve the readout, provided that the nonuniformity of the valley energy levels is small. 
Conversely, when the valley splitting is larger than the Zeeman splitting, the nonuniformity degraded the readout. 
In both cases, we showed that there are regions where the measurement time $t_{\rm meas}$ is much less than the decoherence time $t_{\rm dec}$ such that $t_{\rm dec}/t_{\rm meas}>100$. 
This suggests that less than 1\% measurement error is anticipated, which opens up the possibility for implementing surface codes even in the presence of valley splitting.
\end{abstract}

\maketitle

\section{Introduction}
{\it Background}---
Quantum computers have undergone rapid developments, targeting near-future applications\cite{Google2023,IBM2023,Fowler}. In this respect, silicon qubits have attracted attention because of their affinity to advanced semiconductor circuits, 
which leads to the integration of qubits to build surface codes and control complementary metal-oxide semiconductor (CMOS) circuits on the same chip~\cite{Loss,Zwanenburg}. 
At present, advanced commercial transistors have entered the 2-nm gate-length generation with three-dimensional stacked structures~\cite{TSMC2023,Intel2023,IMEC2023,Kim,Ryckaert}. 
Because quantum effects become more prominent as the device size decreases, 
semiconductor qubits can get the most benefit from advanced semiconductor technologies~\cite{Michniewicz}.

{\it Readout of spin qubits}---The reading out of qubit states is one of the most important process to control the qubit.
The Elzerman readout~\cite{Elzerman}, which is currently a mainstream for silicon qubits,
uses a Zeeman-split spin 1/2 electron in a quantum dot (QD) as the qubit.
The experimental setup of the Elzerman readout consists of a QD, an electrode (reservoir) weakly tunnel-coupled to the QD, 
and a charge sensor that monitors the charge state of the QD~\cite{Koch,Keith}.
The charge sensor detects the spin-dependent charge transfer between the QD and the electrode.
However, this setup of the electrodes and the charge sensor for each qubit requires a large circuit area, 
when the qubits are integrated.
In addition, each electrode and charge sensor requires multiple wiring, 
so the wiring circuitry is inevitably complex.
These are factors that limit the large-scale integration of qubits with the Elzerman readout.

Regarding this readout process, we proposed a different method, a resonant-tunneling readout\cite{tanaJAP}. 
This method is superior to the Elzerman readout in terms of the small number of circuit parts required for readout,
 and the simplicity of the circuit wiring, which are suitable for large-scale integration of qubits.
In the resonant-tunneling readout, 
a "channel-QD" is tunnel-coupled to a qubit (Zeeman-split spin 1/2 of a single electron QD).
The channel-QD is also tunnel-coupled to the source and drain electrodes, 
and generates the resonant-tunneling current between the source and drain (Fig.1). 
The qubit is read out by utilizing the fact that the resonant-tunneling current depends on the qubit state ($\upa$ or $\dna$).
In a system in which two QDs are coupled to one channel-QD, 
it is possible to distinguish between the three states of the two qubits: $\dna\dna$, $\dna\upa$ or $\upa\dna$, and $\upa\upa$ (Fig.1(b)). 
Therefore, when qubits and channel-QDs are arranged side by side (Fig.1(a)), 
it is possible to integrate all qubits and readout the results of the qubit states.
In addition to the qubit readout mechanism, when no source-drain voltage is applied, 
the channel-QD also functions as a coupler that mediates the coupling of two adjacent qubits (two qubit operation).
More details are provided below.

{\it Valley-splitting problem}--- 
Bulk silicon conduction band has sixfold degenerated valley states. 
The degenerated valley states are split into the upper four states and lower two states in silicon qubits.
The upper four states can be separated by the sharp interface, 
and the lowers two energy states (which are written as $E_{\rm V+}$ and $E_{\rm V-}$ in the following)
have to be considered in the process of the spin qubit operations. 
It has been pointed out that the nearly degenerate valley degrees of freedom 
have a negative effect on qubit operations~\cite{Sankar,Xuedong}.
Many researchers have utilized these additional degrees of freedom of the valley states
as new qubit states to control 
the spin states~\cite{Goswami,Dzurak,Eriksson1,Hao,Eriksson2,Ferdous,Zhang,Cai,Degli,Smelyanskiy,Culcer,Gamble,Losert,Bosco}.
However, many past proposals, 
including our proposal above, did not appropriately take the valley degree of freedom into account, 
from viewpoint of the integrated qubit system.

\begin{figure}
\centering
\includegraphics[width=8.5cm]{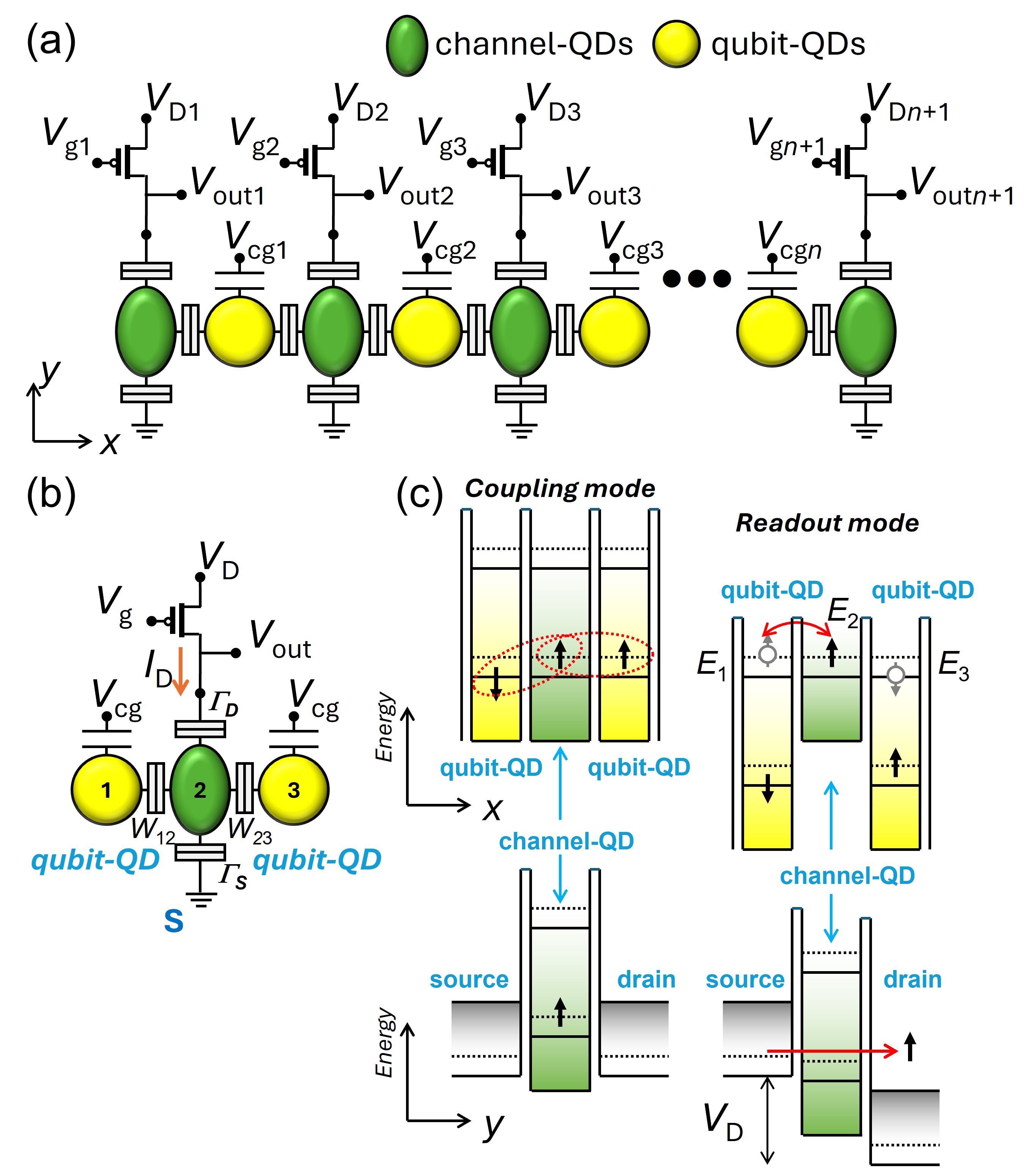}
\caption{
Qubit structure proposed in Ref.~\cite{tanaJAP}, where there were no valley splittings.
(a) Qubit array which consist of the $n$ qubits and transistors.
The qubits (yellow circle) and the channel (green circle) are placed side by side.
A control gate attached to the qubit array controls the electric potentials of the qubits. 
The channel current $I_{\rm D}$ reflects the qubit states.
Static magnetic field $B_z$ is applied.
(b) Single unit of the three-QD system calculated in this study. 
The channel-QD is directly connected to the conventional transistor.
$\Gamma_{\alpha}$ denotes the tunneling couplings between the electrodes ($\alpha=S,D$, $s=\pm 1/2$),
and $W_{ij}$s are the coupling strengths between the QDs.
(c)
The QD band structures for $x$ and $y$ directions of Fig.(b).
The thick arrows express electron spins.
By changing the energy level of the channel-QD, 
two modes are available: coupling ($V_{\rm D}=0$) and readout ($V_{\rm D}\neq 0$) modes. 
The circles denote the energy level to which electrons of channel-QD can tunnel. 
$E_i$ ($i=1,2,3$) represents resonant energy levels of QDs without the magnetic field in the readout mode.
}
\label{fig1}
\end{figure}

{\it Purpose of this study}---
Here, we describe how the existence of the valley splitting
affects the readout process using the resonant-tunneling 
of Ref.~\cite{tanaJAP}.
In order to treat the valley splittings, we newly formulate the transport properties of 
three coupled QDs in which each QD has two energy levels, by using Green function method.
Note that a single energy level is assumed in most 
of the conventional cases~\cite{Jauho,Kim2,tanaPRB}.
By extending the theory to the case of two energy levels in each QD, 
we can treat more general case regarding the three-QD system. 

{\it Outline of this study}---
The rest of this study is organized as follows.
In Section~\ref{sec:model}, our basic model is explained, 
and in Section~\ref{sec:Green} the formalism using 
the Green's function method for the valley splitting cases and transistor models are described. 
In Section~\ref{sec:results}, the numerical results are presented.
In Section~\ref{sec:discussions}, discussions regarding our results are provided.
Section~\ref{sec:conclusions} summarizes and concludes this study.
Appendix presents additional explanations, including a detailed derivation of the Green functions
and a discussion on large valley splitting.

\section{Model}\label{sec:model}
\subsection{Basic structure of the readout using resonant-tunneling}
{\it Device structure---}
Here, we explain the basic spin qubit system without the valley splitting proposed in Ref.~\cite{tanaJAP}. 
We utilize the nonlinear current behavior of the resonant-tunneling of the channel-QD 
to detect the spin states of the qubit-QDs (Fig.1(a)).
Figure~\ref{fig1}(b) is the single unit
which consists of a channel-QD coupled with the qubit-QDs that represents qubits.
The qubit is represented by an electron spin confined in the qubit-QD with its lowest energy level.
The qubit states $|0\ra$ and $|1\ra$ are defined as the $\dna$-spin and $\upa$-spin states, respectively.
The channel-QD coupled to the qubit-QDs is connected to the source and drain electrodes.
The drain electrode is connected to a conventional transistor such as metal-oxide-semiconductor field-effect transistors(MOSFETs), 
which controls the channel current $I_{\rm D}$.
The channel-QD plays both roles of the coupling between the qubits, and the readout.
The electric potential of the qubit-QDs is controlled by the control gates $V_{\rm cg}$.
The electric potential of the channel-QD is also changed by adjusting that of the source and drain voltage $V_{\rm D}$.
Static magnetic field $B_z$ is applied to generate Zeeman splitting.
The both ends of the qubit array are channel-QDs which couple one qubit which corresponds 
$W_{12}=0$ or $W_{23}=0$ in Fig.~\ref{fig1}(b).

{\it Single qubit operation}---
Single qubit operations are carried out in a conventional way 
by using the gradient magnetic field method~\cite{Takeda2,Yoneda}
(local magnets are not shown in Fig.~\ref{fig1}).

{\it Qubit-qubit coupling}---
The coupling between qubits are carried out when there is no applied voltage ($V_{\rm D}=0$).
The electron in the channel-QD mediates the coupling between the 
two qubits (upper-left figure of Fig.1(c)).
The coupling between qubits are 
described as the three QD system which has been 
already investigated in the literature\cite{Fei,Burkard,Kandel,Noiri,Medford}.

{\it Readout process}---
Readout of the qubit states is carried out by applying $V_{\rm D}$. 
In order not to directly change the qubit states,
the energy level of the qubit-QD(QD${}_1$ or QD${}_3$) is lowered downward
as shown in the upper-right figure of Fig.1(c).
This situation is realized by applying $V_{\rm cg}$.
By adjusting the energy level of the channel-QD(QD${}_2$) to the upper energy-level of the singlet states
of the qubits, the channel electrons go back and forth between these energy levels 
and reflect the qubit states on $I_{\rm D}$, as denoted by the red arrows in the right figures of Fig.1(c).

It is also assumed that the qubit-QD has a large on-site Coulomb energy $U$~\cite{Engel}.
The spin direction of the upper energy levels of the singlet is 
opposite to that of the lower electron spin.
The $I_{\rm D}$ changes 
when the energy level of the channel-QD resonates with those of the upper energy levels of the singlet states.
The backaction of the readout is estimated 
by comparing the measurement time $t_{\rm meas}$ with a coherence time $t_{\rm dec}$.
The ratio $t_{\rm dec}/t_{\rm meas}$ corresponds to the 
possible number of the readout, which is desirable to exceed more than 
hundred to realize the surface code.
In Ref.~\cite{tanaJAP}, we found the parameter regions in which $t_{\rm dec}/t_{\rm meas}>100$ is satisfied.

Under the applied magnetic field, both the energy levels of the QDs and the source and drain exhibit spin-splitting~\cite{Sanchez}, 
as illustrated later in Fig.~\ref{figRT}.
We assume that the up-spin current ($\upa$-current) and the down-spin current ($\dna$-current) are independent, 
then, the $\upa$-current and $\dna$-current can be treated to have 
different Fermi energies $E_{F\pm}=E_F \mp \Delta_z/2$.
The $\upa$ and $\dna$-currents are controlled by the transistor
which is connected to the channel-QD.

Stacking the three-QD array system will form the surface code, 
if the forthcoming 2nm-transistor architecture~\cite{TSMC2023,Intel2023,IMEC2023,Kim,Ryckaert} is used.
Two types of stacking forms are discussed in appendix~\ref{app:array}.

\subsection{Two regions of valley splitting}
The valley-splitting energy $E_{\rm VS}\equiv E_{V+}-E_{V-}$ between the two valley states $E_{V+}$ and $E_{V-}$ 
are approximately within the range from 10 $\mu$eV to 2 meV\cite{Goswami,Gamble}.
The valley splitting can be treated separately into three regions 
relating with an applied magnetic field described by the Zeeman splitting energy $\Delta_z\equiv g\mu_B B_z$:\\
(1) $E_{\rm VS}<\Delta_z$ (small valley region), \\
(2) $E_{\rm VS}\approx \Delta_z$ (intermediate valley region), \\ 
(3) $E_{\rm VS}>\Delta_z$ (large valley region). \\
In this study, the small valley region of $E_{\rm VS}<\Delta_z$ is mainly considered.
The large valley region of $E_{\rm VS}>\Delta_z$ is treated in appendix~\ref{app:large_valley}.
In these two regions,  no spin flip can be assumed.
The intermediate region $E_{\rm VS}\approx\Delta_z$, 
where the spin flips happen, is not treated here.

In general, the valley energies are randomly distributed in space due to variations in the fabrication process. 
Consequently, it is possible for the three quantum dots (QDs) depicted in Fig. 1(b) to have different valley energies, 
indicating a nonuniformity in these energies. In the latter part of this study, 
we will present the current characteristics for different energy levels among the QDs. 
The valley splitting is modeled by replacing a single energy level $E_i$ in a QD with two energy levels 
$E_{ia}$ and $E_{ib}$ ($i=1,2,3)$. 
The source and drain electrodes consist of three-dimensional electrons, 
and it is assumed that there are no valley splittings in the electrodes, similar to bulk silicon.


\subsection{Model for spin qubit system with small valley splitting}
\begin{figure}
\centering
\includegraphics[width=8.5cm]{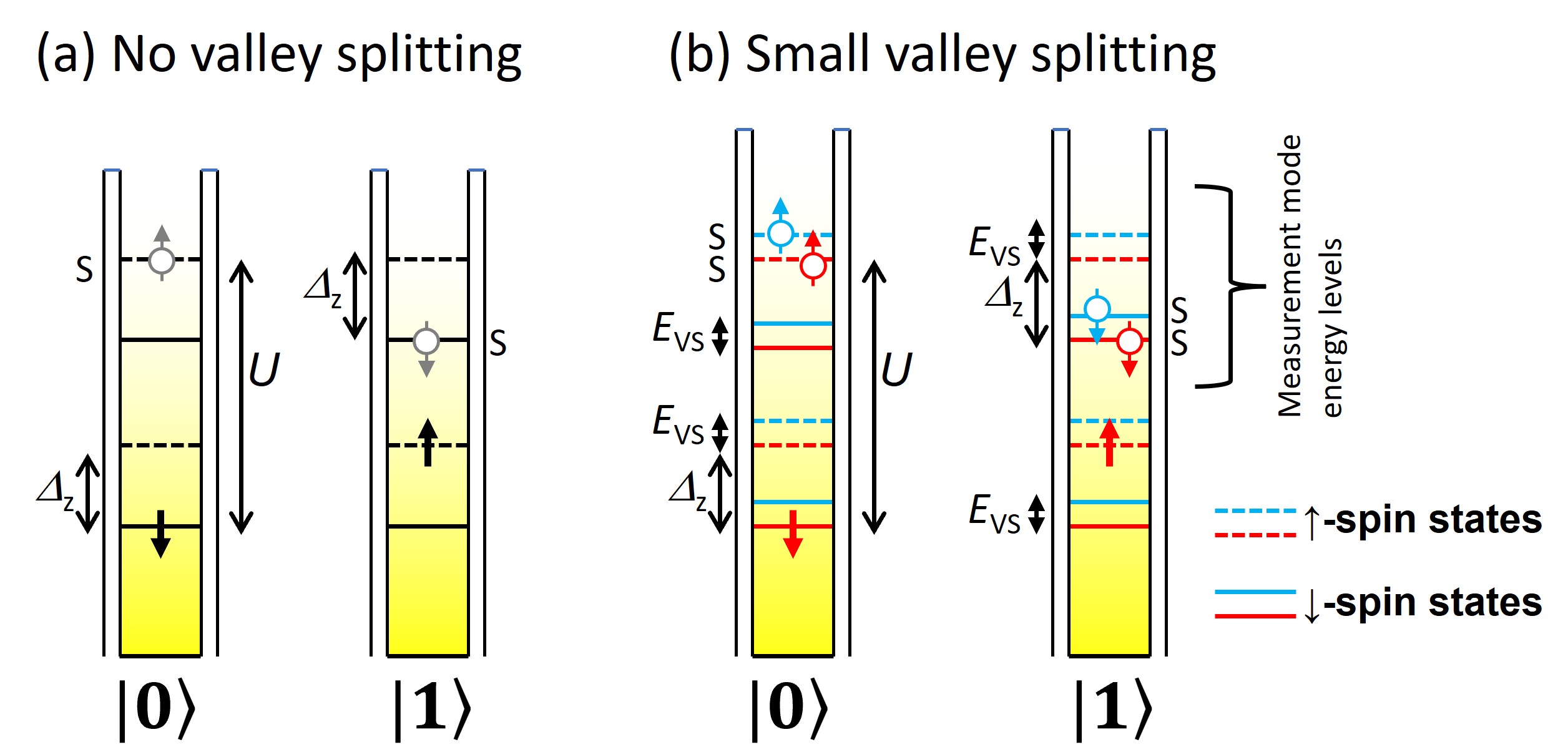}
\caption{
The energy band diagram of the two qubit states ($|0\ra$ or $|1\ra$). 
(a) No valley splitting case. (b) Small valley splitting case of $E_{\rm VS}<\Delta_z$. 
The solid arrows indicate the spins of the qubits (first electron). 
The arrows with white circles indicate the possible spin state which the second electron 
enters the QDs from the channel-QD. 
The second electron can enter the QD forming a singlet state ($S$) 
whose energy level exist $U$ higher than the lowest energy level of the QD.
}
\label{fig2}
\end{figure}

Figure~\ref{fig2} shows the change in the energy band diagram when there are small valley splittings for qubit-QDs ($E_{\rm VS}<\Delta_z$). 
Because of the valley splitting, the energy diagram of qubit-QDs changes from Fig.~\ref{fig2}(a) to Fig.~\ref{fig2}(b). 
A large on-site Coulomb energy $U$ is assumed ($U\gg \Delta_z, E_{\rm VS}$).

\begin{figure}
\centering
\includegraphics[width=7.5cm]{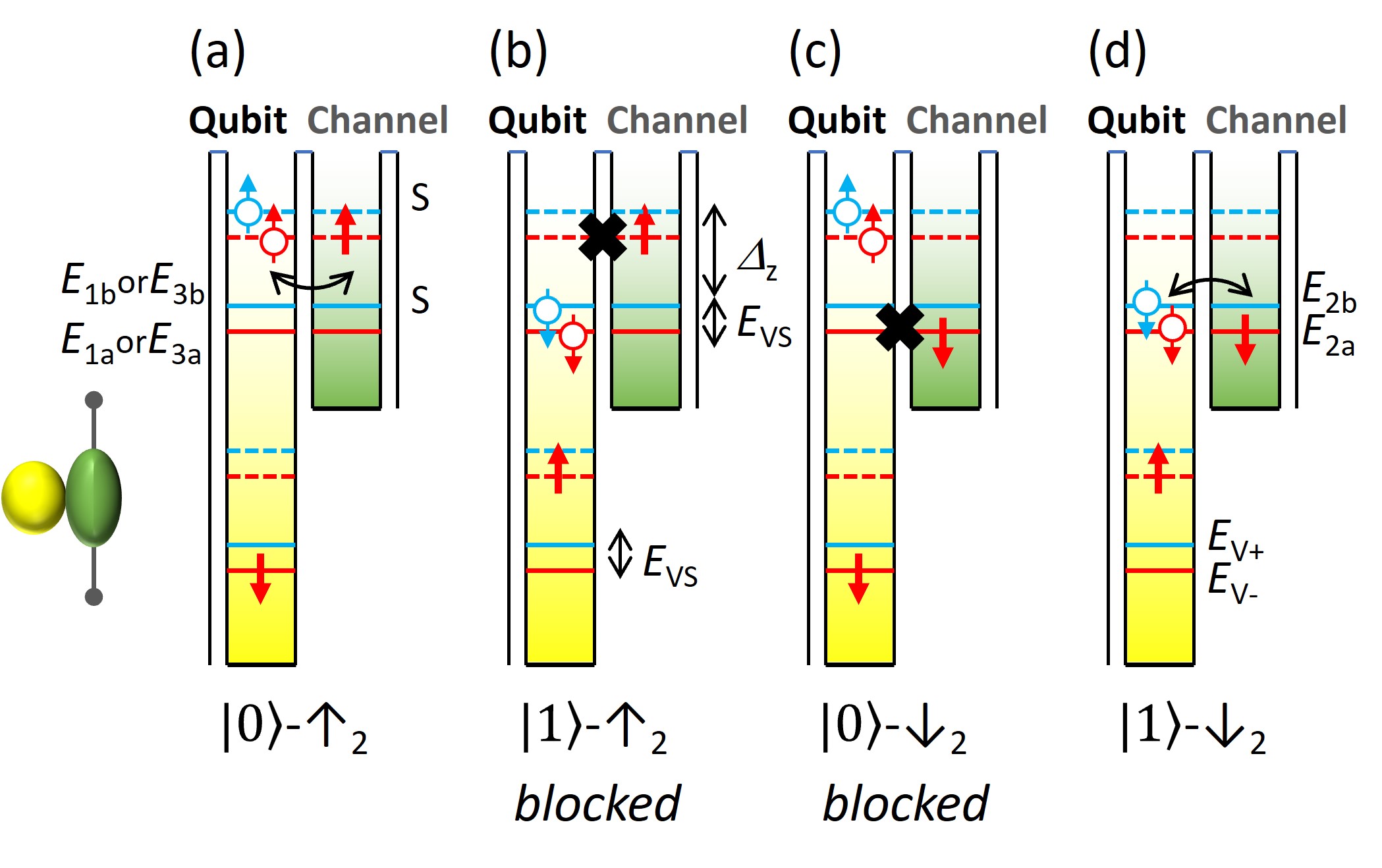}
\caption{
The tunneling profile between the channel-QD(QD${}_2$) and qubit-QD for small valley splitting.
Here, the qubit-QD represents both QD${}_1$ and QD${}_3$.
(a) and (b) for the $\upa$-current. (c) and (d) for $\dna$-current.
From (a) and (b), $\upa$-current can exchange $\upa$-spin electron only when the qubit state is $|0\ra$.
From (c) and (d), $\dna$-current can exchange $\dna$-spin electron only when the qubit state is $|1\ra$. 
$E_{V-}$ and $E_{V+}$ are the valley energies of the qubit.
$E_{ia}$ and $E_{ib}$ are the valley energies of the singlet state, 
where $E_{\rm VS}=E_{V+}-E_{V-}=E_{ia}-E_{ib}$ ($i=1,2,3)$.
}
\label{fig3}
\end{figure}

Figure~\ref{fig3} shows the energy diagram of the relationship between the qubit-QDs (QD${}_1$ or QD${}_3$) and the channel-QD (QD${}_2$). 
In the readout mode, the bottom of the energy band of the channel-QD is higher than those of the qubits (Fig.~\ref{fig1}(c)). 
Depending on the qubit states ($|0\ra$ or  $|1\ra$) and currents ($\upa$ or $\dna$-currents), 
different distributions of the current characteristics are considered. 
The readout mechanism is described such that if there is an energy level in the qubit-QDs (circles in Figs.~\ref{fig2} and \ref{fig3} for the qubit-QDs), 
electrons with the same spin can tunnel into the qubit-QDs. When there is no corresponding energy for a given $V_{\rm D}$, the tunneling is blocked. 
This blocked feature is expressed by the no tunneling term in the Hamiltonian shown below ($W_{i\xi,j\xi'}=0$ in Eq.(\ref{Hamilt})). 

From Fig.~\ref{fig3}, 
the $\upa$-spin electron of the $\upa$-current can enter the qubit-QD only when the qubit-QD is in $|0\ra$, and $\dna$-spin electron of the $\dna$-current can enter the qubit-QD only when the qubit-QD is in $|1\ra$.
Depending on the spin states of the two-qubit-QD (QD1 or QD3), four qubit states $|00\ra$(or $|\dna\dna\ra$), $|01\ra$(or $|\dna\upa\ra$), $|10\ra$(or $|\upa\dna\ra$), and $|11\ra$ (or $|\upa\upa\ra$) exist. 
Because $|\dna\upa\ra$ has the same effect as $|\upa\dna\ra$, only $|\upa\dna\ra$ is considered in the following.

The valley splitting energy of $E_{\rm VS} \ll 100\mu$eV is close to the expected operating temperature of approximately 100 mK (which is around 10 $\mu$eV). 
As a result, the mixing of valley energy levels ($E_{ia}$ and $E_{ib}$($i=1,2,3$)) during the tunneling process must be taken into account,
regardless of whether the qubit electron (the first electron) is positioned at the lower ($E_{V-}$) or the upper ($E_{V+}$) of the valley energy level.

\section{Formulation}~\label{sec:Formulation}
\subsection{Green function methods}~\label{sec:Green}
In the readout process, the energy levels $E_{ia}$ and $E_{ib}$ ($i=1,2,3)$  
are considered as depicted in Fig.~\ref{fig3}.
Thus, we formulate three QDs each of which have two valley energy levels.
Strong uniform magnetic field $B_z$ is applied, and there is no gradient magnetic field.
Because no spin flip is assumed, 
the $\upa$ and $\dna$ spins can be treated separately.
Then, the Hamiltonian of the three QDs depicted in Fig.~\ref{fig3} is given by
\begin{eqnarray}
H&=&\sum_{i=1}^3  \sum_{s=\upa,\dna} \sum_{\xi=a,b}E_{i\xi s} d_{i\xi s}^\dagger d_{i\xi s}  
\!+\! \!\sum_{\alpha=S,D}\sum_{k_\alpha,s} E_{k_\alpha s} f_{k_\alpha s}^\dagger f_{k_\alpha s}
\nonumber \\ 
&+&\!\sum_{\alpha=S,D}\sum_{k_\alpha,s}\sum_{\xi=a,b} 
[V_{k_\alpha,s,\xi} f_{k_\alpha,s}^\dagger d_{2\xi s}+V_{k_\alpha s}^* d_{2\xi s}^\dagger f_{k_\alpha s} ]
\nonumber \\
&+& \sum_{i=1}^{2}\sum_{s=\upa,\dna}  \sum_{\xi,\xi'=a,b} W_{i\xi,i+1\xi'}(d_{i\xi s}^\dagger d_{i+1,\xi',s}+h.c.),
\label{Hamilt}
\end{eqnarray}
where $f_{k_\alpha,s}^\dagger$ ($f_{k_\alpha,s}$) creates (annihilates) an electron of momentum $k_\alpha$ 
and spin $s(=\pm 1/2)$ in the electrodes $(\alpha=S,D)$. 
$d^\dagger_{i\xi s}$ ($d_{i\xi s}$) creates (annihilates) 
an electron in the QDs ($i=1,2,3$) where $\xi=a$ and $\xi=b$ correspond 
to the lower and higher valley levels. 
$E_{k_\alpha s}=E_{k_\alpha}+s g\mu_B B_z$ is the energy level of the electrode ($\alpha=S,D$), 
and $E_{i\xi s}=E_{i}+sg\mu_B B_z+\Xi_\xi$ is the energy level for three QDs ($i=1,2,3$)
with $\Xi_a=-E_{\rm VS}/2$ and $\Xi_b=E_{\rm VS}/2$. 
$E_{i}$ is an energy level without the valley splitting or magnetic fields.
 
The coupling coefficients of the electrodes to the channel-QD are given by
\begin{equation}
\Gamma_{\alpha,s,\xi}(\omega)=2\pi \sum_{k_\alpha} |V_{k_\alpha s,\xi}|^2\delta (\omega-E_{k_\alpha s}),
\label{Gamma}
\end{equation}
($\alpha=S,D$, $\xi=a,b$).
Strictly speaking, $\Gamma_{\alpha,s,\xi}(\omega)$ depends on the spin direction 
through the density of states, and the valley levels. 
However, for the sake of simplicity, we take 
$\Gamma_\alpha\equiv \Gamma_{\alpha,\uparrow,\xi}(\omega)=\Gamma_{\alpha,\downarrow,\xi}(\omega)
(\alpha=S,D,\xi=a,b)$.

Following Ref.~\cite{Jauho,Meir1}, the current $I_{S}$ of the source electrode 
is derived from the time-derivative of the number of the electrons 
$N_{S}\equiv \sum_{k_Ss}f_{k_Ss}^\dagger f_{k_Ss}$ 
of the source electrode given by
\begin{eqnarray}
I_{S}(t)&=&-e\left\langle \frac{d N_{S}}{dt}\right\rangle 
=-\frac{ie}{\hbar} \langle [H,N_{S} ]\rangle \nnm \\
&=&\frac{ie}{\hbar} \sum_{k,s}\sum_{\xi=a,b}[V_{k_Ss}\langle f_{k_S s}^\dagger d_{2\xi s} \rangle
-V_{k_Ss}^* \langle d_{2\xi s}^\dagger f_{k_Ss} \rangle] \nnm \\
&=& \frac{2e}{\hbar} \int \frac{d\omega}{2\pi} {\rm Re} \left\{
\sum_{ks}\sum_{\xi=a,b} V_{k_Ss} G^<_{d_{2 \xi s},f_{k_S s}}(\omega)
\right\}, 
\label{currentformula}
\end{eqnarray}
where
\begin{eqnarray}
G^<_{d_{2\xi s},f_{k_\alpha s}}(t,t')&\equiv& i \langle f_{k_\alpha s}^\dagger(t') d_{2\xi s}(t)\rangle, \\
G^>_{f_{k_\alpha s}d_{2\xi s}}(t,t')&\equiv& -i \langle f_{k_\alpha s}(t) d_{2\xi s}^\dagger(t')\rangle, 
\end{eqnarray}
and
\begin{eqnarray}
G^<_{f_{k_\alpha s},d_{2\xi s}}(t,t)=-[G^<_{d_{2\xi s},f_{k_\alpha s}}(t,t)]^*.
\end{eqnarray}
When the direction of the drain current $I_{\rm D}$ is defined as the increase in the number of electrons in the drain electrode, we can take $I_{\rm S}=I_{\rm D}$.
Because we assume that the spin-flip process is neglected, the suffix $s$ is omitted in the following.

The Green functions are derived using the equation of motion method~\cite{Jauho}.
For example, the time-dependent behavior of the operator $f_{k_\alpha,s}$ is derived from $i \frac{df_{k_\alpha,s}}{dt}=[H,f_{k_\alpha,s}]$, and we have
\begin{eqnarray}
\omega f_{k_\alpha,s} = \frac{i}{\hbar} [H, f_{k_\alpha,s}].
\end{eqnarray}

The Green functions of the electrodes ($\alpha=S,D$) are the free-particle Green functions given by
\begin{eqnarray}
g_\alpha ^<(k,\omega)&=& 2\pi i f_\alpha (E_{k_\alpha}) \delta(\omega-E_{k_\alpha}), \\
g_\alpha ^>(k,\omega)&=& 2\pi i (f_\alpha(E_{k_\alpha})-1) \delta(\omega-E_{k_\alpha}), \\
g_\alpha^r(k,\omega)&=& \frac{1}{\omega-E_{k_\alpha} +i\delta},
\end{eqnarray}
where 
$f_\alpha(\omega)= [\exp[(\omega-\mu_\alpha)/(k_BT)]+1]^{-1}$ ($k_B$, $\mu_\alpha$, and $T$ denote
the Boltzmann constant, the chemical potential of the $\alpha$-electrode,
and temperature, respectively).
As shown in the appendix~\ref{app:green_function}, all Green functions are obtained after a long derivation process.
Eventually, the current formula is expressed by
\begin{eqnarray}
I_{\rm D}\!
&=&\! \frac{e}{h} \!\int \!\! d\omega \!\!
\left[P\frac{1}{\omega\!-\! E_{2a}}\!+\! P\frac{1}{\omega\!-\! E_{2b}}\right]^2
\!\! \frac{
\Gamma_S\Gamma_D(f_S(\omega)\!-\! f_D(\omega))
}{|b_a^r b_b^r- a_{vba}^r a_{vab}^r|^2 },
\nnm \\
\label{current_formula}
\end{eqnarray}
where $b_a^r$, $b^r$, $a_{vba}^r$, and $a_{vab}^r$ are the retarded expressions of the functions given by
\begin{eqnarray}
b_a&=& 1- (\Sigma_{f}+A_{va})\Sigma_{2a}, \nnm\\
b_b&=& 1- (\Sigma_{f}+A_{vb})\Sigma_{2b}, \nnm\\
a_{vab}&=& (\Sigma_{f}+A_{va})\Sigma_{2a}, \nnm\\
a_{vba}&=& (\Sigma_{f}+A_{vb})\Sigma_{2b}.
\end{eqnarray}
Here, we define
\begin{eqnarray}
\Sigma_f^<&=&
\sum_{k_\alpha} \left( \frac{|V_{k_\alpha,s,\xi}|^2}{\omega-E_{k_\alpha}} \right)^<
=i[\Gamma_S f_S(\omega)+\Gamma_D f_D(\omega)],
\nnm \\
\Sigma_f^r&=&
\sum_{k_\alpha} |V_{k_\alpha,s,\xi}|^2 \left( P\frac{1}{\omega-E_{k_\alpha}}-i\pi \delta(\omega-E_{k_\alpha}) \right)
\nnm \\
& =&\Lambda_f(\omega) -i\gamma_f,
\end{eqnarray}
with 
$\gamma_f\equiv [\Gamma_S +\Gamma_D ]/2$
($\Lambda_i(\omega)$ is assumed to be constant and included in $E_i$ in the following),
and 
$\Sigma_{2a}\equiv 1/(\omega-E_{2a})$
and $\Sigma_{2b}\equiv 1/(\omega-E_{2b})$.
$A_{va}$ and $A_{vb}$ are given by
\begin{eqnarray}
A_{va}&\equiv&
|W_{12aa}|^2 \Sigma_{1a}
+|W_{12ab}|^2 \Sigma_{1b}
\nnm \\
& &+|W_{32aa}|^2 \Sigma_{3a} 
+|W_{32ab}|^2 \Sigma_{3b},
\nnm \\
A_{vb}&\equiv &
|W_{12bb}|^2 \Sigma_{1b}+ |W_{12ba}|^2 \Sigma_{1a}
\nnm  \\
& &+|W_{32bb}|^2 \Sigma_{3b}+ |W_{32bb}|^2 \Sigma_{3a},
\end{eqnarray}
where
$\Sigma_{1\xi}\equiv 1/(\omega-E_{1\xi})$ and $\Sigma_{3\xi}\equiv 1/(\omega-E_{3\xi})$ ($\xi=a,b$).
$W_{12\xi\xi'}$ expresses the coupling between the $1\xi$ state and the $2\xi'$ state. We assume that $W_{12ab}=W_{12aa}$, $W_{32ab}=W_{32aa}$, $W_{12ba}=W_{12bb}$, and $W_{32ba}=W_{32bb}$, 
which means that the tunneling probabilities between different valley energies are the same as those between the same valley energies.
Using these formalisms, four distributions can be distinguished as follows.
For only the channel-QD without coupling to qubits (no qubit case), $W_{12aa\upa}=W_{12aa\dna}=W_{32aa\upa}=W_{32aa\dna}=0$ is taken for the $\upa$ and $\dna$-currents.
For $|\upa\dna\ra$, $W_{12aa\upa}=W_{32aa\dna}=W$ and $W_{12aa\dna}=W_{32aa\upa}=0$ are taken 
(for the state $|\dna\upa\ra$, the symbols $\upa$ and $\dna$ are exchanged).
For the $|\upa\upa\ra$ case, $W_{12aa\upa}=W_{32aa\upa}=W$ for the $\upa$-current and $W_{12aa\dna}=W_{32aa\dna}=0$ for the $\dna$-current are taken.
For the $|\dna\dna\ra$ case, $W_{12aa\upa}=W_{32aa\upa}=0$ for the $\upa$-current and $W_{12aa\dna}=W_{32aa\dna}=W$ for the $\dna$-current are taken.

In the large valley region, $E_{\rm VS}>\Delta_z$ and $E_{\rm VS}>V_{\rm D}$ are assumed, and the effect of the valley splitting is neglected. 
This is the same situation as that in our previous paper~\cite{tanaJAP}. 
The current $I_{\rm D}$ for the large valley splitting case is given by
\begin{eqnarray}
I_{\rm D}
&=& \frac{e}{h} \int \! d\omega \!
\left[P\frac{1}{\omega-E_{2a}}\!\right]^2
 \frac{
\Gamma_S\Gamma_D(f_S(\omega)-f_D(\omega))
}{|b_b^r|^2 }.
\nnm \\
\end{eqnarray}

\subsection{Transistor model}
We use the core model for $I_{\rm D}$-$V_{\rm D}$ of the fin field-effect transistor (FinFET), given by~\cite{BSIM}
\begin{equation}
I_{\rm TR} = \beta (V_{\rm g}-V_{\rm th}-V_{\rm ds}/2)V_{\rm ds}, 
\label{CMOS}
\end{equation}
where $\beta\equiv (\mu {\mathcal W}/L) (\epsilon/EOT)$ ($L=1$~{\textmu}m, ${\mathcal W}=80$~nm, $\mu=1000$~cm$^{2}$V$^{-1}$s$^{-1}$, $EOT=1$~nm, and $\epsilon=3.9\times 8.854\times 10^{-12}$~F/m represent the gate length, 
gate width, mobility, oxide thickness, and dielectric constant, respectively~\cite{BSIM}).
The output voltage $V_{\rm out}$ is numerically determined by solving the equations of the QD device and the transistor, given by
\begin{equation}
V_{\rm D}=V_{\rm out}+V_{\rm ds}.
\label{eq1out}
\end{equation}
where $V_{\rm ds}$ is the bias difference of the source and drain of the transistor.
This equation is solved numerically using Newton’s method, assuming $I_S=I_{\rm TR}$.

\subsection{Measurement model}
Using the current difference $\Delta I$ depending on different qubit states in Fig.~\ref{fig3}, 
the qubit states are distinguished. The measurement time $t_{\rm meas}$ is estimated using $\Delta I$ and shot noise $S_N$, given by~\cite{Schon}
\begin{equation}
t_{\rm meas}^{-1}\equiv \frac{(\Delta I)^2}{4S_N}.
\label{measurement}
\end{equation}
The measurement times of the four cases in Fig.~\ref{fig3} are calculated from the current difference $\Delta I$ from the reference state where there is no coupling to the qubit (no qubit case).
Here, the classical form $S_N=2eI_{\rm D}$ of the shot noise is used.
 
Although the decoherence time was calculated using Fermi’s golden rule previously~\cite{tanaJAP}, 
in this study, we introduce a fixed coherence times $t_{\rm dec}$.
This is because it is hard to disassemble the many tunneling processes in the present model, 
and it is rather beneficial to use a fixed coherence time to directly compare the theory to experiments.

\begin{figure}
\centering
\includegraphics[width=8.5cm]{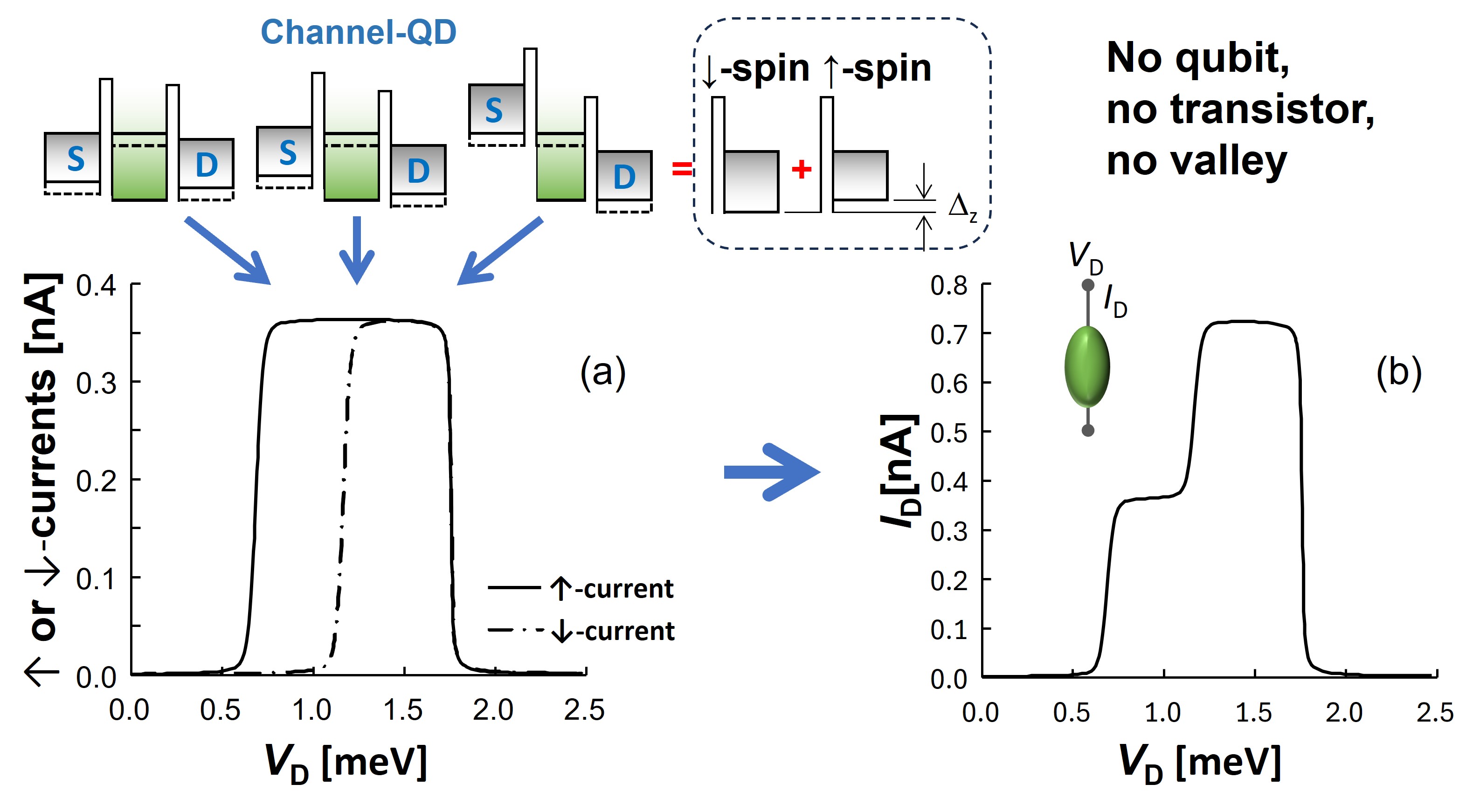}
\caption{
$I_{\rm D}$-$V_{\rm D}$ characteristics of a simple resonant-tunneling consisting of a channel-QD with the source and drain under the magnetic field (Zeeman splitting is given by $\Delta_z=0.232$ meV).
There is neither qubit, transistor, nor valley energies. 
(a) The $\dna$-spin and $\upa$-spin currents are depicted separately.
The switching-on voltages are different between the $\dna$-spin and $\upa$-spin currents. 
However, the resonant peaks end at the same $V_{\rm D}$.
$\Gamma_0=0.03u_0=3.0\times10^{-6}$ eV ($u_0=10^{-4}$ eV), $E_F$=1 meV, $T=100$~mK,
and $E_2^{(0)}=0.8(E_F+u_0)=0.88$~meV. 
(b) Total current $I_{\rm D}$ of the $\dna$-spin and $\upa$-spin currents.
}
\label{figRT}
\end{figure}
\begin{figure}
\centering
\includegraphics[width=5.5cm]{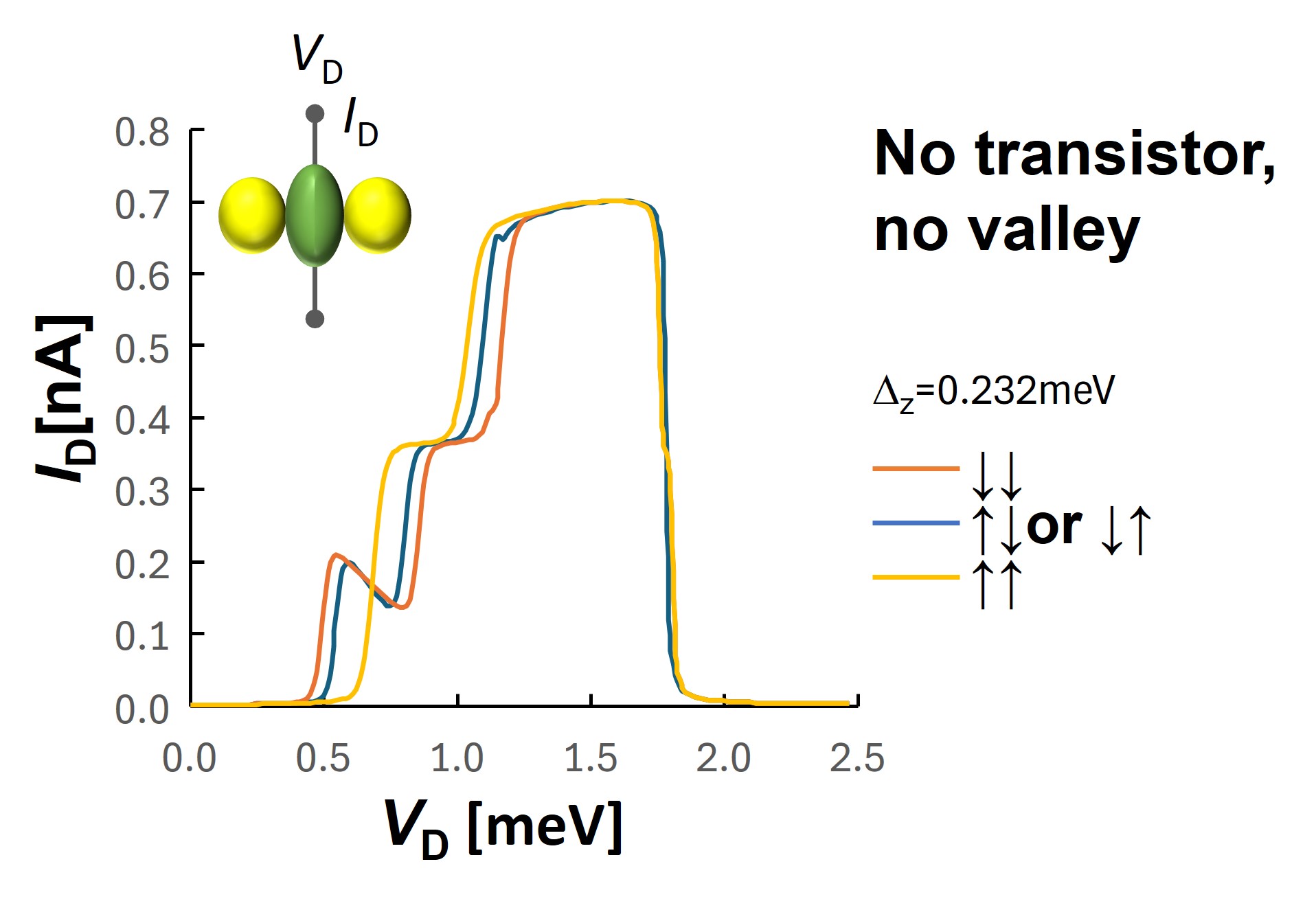}
\caption{
$I_{\rm D}$-$V_{\rm D}$ characteristics coupled with two qubits.
There is neither a transistor nor valley splitting. 
$\Gamma_0=3.0\times10^{-6}$ eV, $E_F$=1~meV, $T=100$~mK.
$E_1=E_3=E_F+2u_0=1.2$ meV, $E_2^{(0)}=0.88$~meV, and $W=0.09$~meV.
}
\label{figNoTrwithqubit}
\end{figure}

\section{Results}~\label{sec:results}
Because the structure is a little complicated, we explain the transport properties step by step.
In Section \ref{RT1}, a simple resonant-tunneling structure is explained, where there is neither qubit, transistor, nor valley. 
In Section \ref{RT2}, the $I_{\rm D}$-$V_{\rm D}$ characteristics of the channel-QD with qubits are discussed, where there is neither transistor nor valley. 
In Section \ref{RT3}, the $I_{\rm D}$-$V_{\rm D}$ characteristics of the channel-QD with valley splitting but without qubits or a transistor are explained. 
In Section \ref{uniform}, the $I_{\rm D}$-$V_{\rm D}$ characteristics with uniform valley splittings are shown depending on the qubit states with the transistor. 
In Section \ref{nonuniform}, the general $I_{\rm D}$-$V_{\rm D}$ characteristics with different valley splitting energies in the two qubit-QDs are explained. 
The numerical results of the output voltage $V_{\rm out}$ are also presented. 
Finally, in Section \ref{time}, the number of possible measurements $t_{\rm dec}/t_{\rm meas}$ are shown.
The results of the large valley spitting cases are shown in the appendix~\ref{app:large_valley}.

\subsection{Basic resonant-tunneling properties without qubits, transistor, or valley splitting}\label{RT1}
Figure~\ref{figRT} shows the current characteristics of a single QD coupled to the source and drain under a magnetic field. 
Figure~\ref{figRT}(a) shows the  $\upa$ and $\dna$-currents separately. Single peaks are observed for each current element. 
The electrons of the $\upa$ and $\dna$-currents have different energy bands, 
as shown in the upper insets of Fig.~\ref{figRT}(a).
The switching-on voltage of the $\upa$-current is lower than that of the $\dna$-current for this parameter region because the energy level of the $\upa$-spin is higher than that of the $\dna$-current 
(see upper-left inset of Fig.~\ref{figRT}(a)).
Meanwhile, the resonant peaks end at the same $V_{\rm D}$ for the $\upa$, and $\dna$-currents (see middle-right inset of Fig.~\ref{figRT}(a)). 
Figure~\ref{figRT}(b) shows the total current $I_{\rm D}$ without the transistor or valley splitting, which is the summation of the $\upa$ and $\dna$-currents. 
The step structure is the result of the different switching-on voltages of the $\upa$ and $\dna$-currents.

The characteristics of the resonant peak can be approximately analyzed by comparing the energy level $E_2^{(0)}\pm \Delta_z/2$ of channel-QD with those of the two electrodes. 
The detailed analysis for the resonant peak structure is presented in appendix~\ref{analysis_of_rt}. 
Let us shortly analyze Fig.~\ref{figRT} using Table~\ref{tbl2} of appendix~\ref{analysis_of_rt}. 
Because $E_2^{(0)}+\Delta_z/2=0.88+0.116=0.996$~meV is below $E_F$=1~meV, the top row of Table~\ref{tbl2} is applied. 
Then, the switching-on voltages are approximately given by $V_{{\rm D}\upa}^{\rm on}\approx 2(E_F\!-\! E_2^{(0)}\!-\Delta_z/2)=0.008$~meV and $V_{{\rm D}\dna}^{\rm on}\approx 2(E_F\!-\! E_2^{(0)}\!+\Delta_z/2)=0.472$~meV for the $\upa$ and $\dna$-currents, respectively. 
Meanwhile, the switching-off voltages are given by $V_{{\rm D}\dna}^{\rm off}\approx 2E_2^{(0)}=1.76$~meV. 
The width of the resonant peaks $4E_2^{(0)}+\Delta_z-2E_F$ are 1.752 and 1.288~meV, respectively. The peak centers $E_F-\Delta_z/2$ are 0.884 and 1.116~meV. 
If we compare these values with those in Fig.~\ref{figRT}(a), the $V_{{\rm D}\dna}^{\rm off}$ is approximately correct. 
However, the other values are shifted. 
Thus, it is better to use the analysis for obtaining general trends.

\subsection{$I_{\rm D}$-$V_{\rm D}$ characteristics without a transistor or valley splitting}\label{RT2}
Figure~\ref{figNoTrwithqubit} shows $I_{\rm D}$-$V_{\rm D}$ characteristics when two qubits are added to the simple resonant-tunneling structure of Fig.~\ref{figRT}(b). 
Apparently, the current for the pair $\upa\dna$ or $\dna\upa$ exists in the middle of the $\upa\upa$ and $\dna\dna$ pair. 
In the present case, as shown in Fig.~\ref{figRT}, 
the switching-on voltage of the $\upa$-current is lower than that of the $\dna$-current. 
For the $\dna\dna$ pair ($|00\ra$ state), the resonant energy level of the qubits, 
which matches the energy level of the channel-QD, becomes high, as shown in Figs.~\ref{fig3}(a) and \ref{fig3}(c). 
The resonance occurs around $E_1(=E_3)=E_2^{(0)}+V_{\rm D}/2$, which leads to 
$V_{\rm D}\approx 2(E_1-E_2^{(0)})=2(1.2-0.88)=0.64$meV.
The small peak structure of $\dna\dna$ pair around $V_{\rm D}\approx 0.6$ meV is the result of this resonance.
Meanwhile, for the $\dna\dna$ pair ($|11\ra$ state), the resonant energy level of the qubit-QD and the channel-QDs
becomes low, as shown in Figs.~\ref{fig3}(b) and \ref{fig3}(d). 
Thus, the corresponding resonance peak is hidden in the original resonant-tunneling peak of Fig.~\ref{figRT}(b).

\subsection{Resonant-tunneling properties under valley splitting without qubits or a transistor}\label{RT3}
\begin{figure}
\centering
\includegraphics[width=8.5cm]{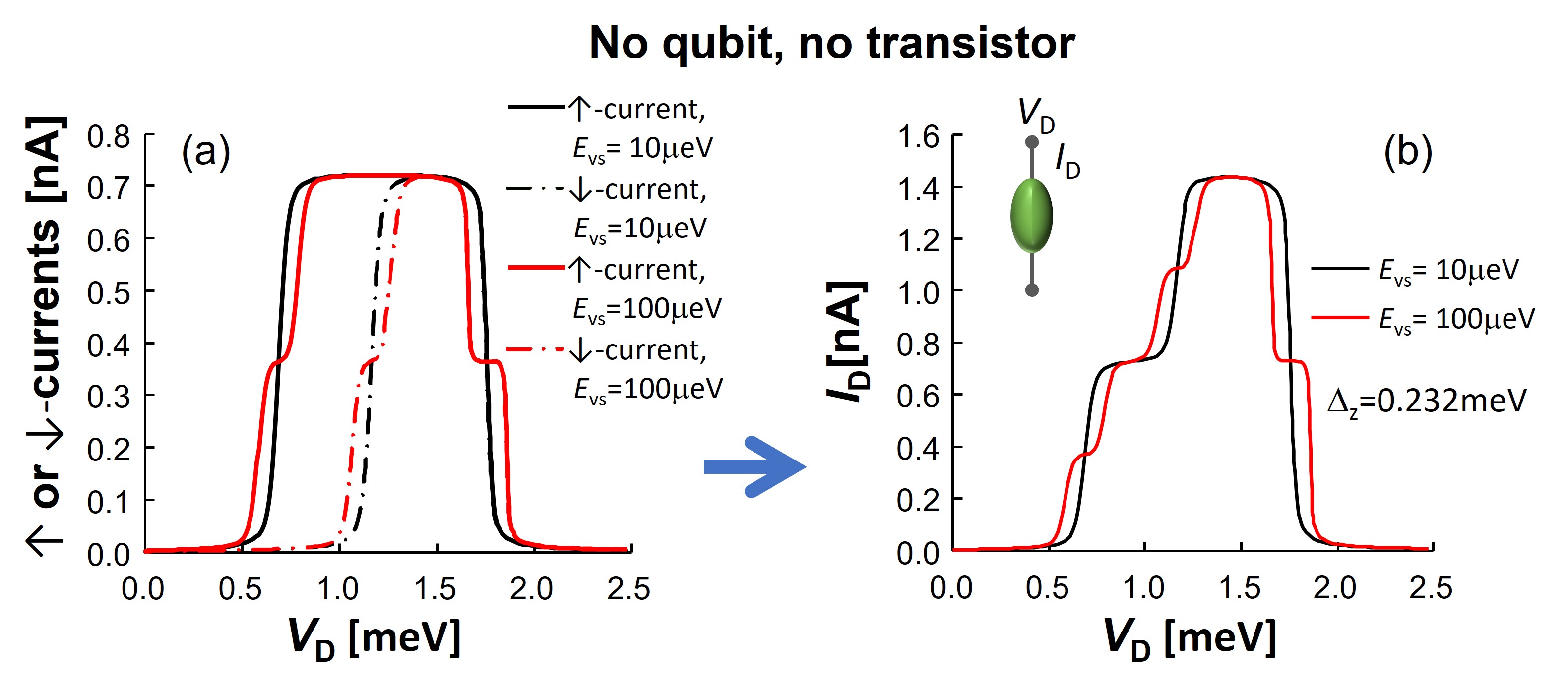}
\caption{
Spin-dependent currents as a function of $V_{\rm D}$ without coupling to qubits ($W=0$) or a transistor. 
(a) $\upa$- and $\dna$-currents are described separately. 
(b) Total current $I_{\rm D}$ consisting of the $\upa$- and $\dna$-currents. 
$\Gamma_0=3.0\times 10^{-6}$ eV, $E_F$=1 meV, $T=100$~mK and $E_2^{(0)}=0.88$~meV. 
The different characteristics between the $\upa$- and $\dna$-currents come from their different Fermi energies (see Fig.~\ref{figRT}). 
}
\label{fignoqubit}
\end{figure}

In Fig.~\ref{fignoqubit}(a), the $\upa$- and $\dna$-currents are separately shown as functions of $V_{\rm D}$ for two valley splitting energies ($E_{\rm VS}=10\mu$eV and $E_{\rm VS}=100\mu$eV). 
The single-peak structures represent the resonant peak by the resonant energy level similar to Fig.~\ref{figRT}. 
In Fig.~\ref{fignoqubit}(b), the total current $I_{\rm D}$ by the $\upa$- and $\dna$-currents are shown. 
A shoulder structure becomes clear as $E_{\rm VS}$ becomes larger.

The shoulder structure is analyzed in Table~\ref{tbl2} of appendix~\ref{analysis_of_rt}. 
The total current is the summation of the currents of $E_{2a}$ and $E_{2b}$ from Eq.(\ref{currentformula}). 
From Table~\ref{tbl2}, the widths of the resonant peak and peak center are approximately estimated by $4E_2^{(0)}+\Delta_z-2E_F$ and $E_F-\Delta_z/2$, respectively. 
If we apply $E_{2a}$ and $E_{2b}$ into $E_2^{(0)}$, the center of the resonant peak does not change, 
but the width of the resonant peak depends on the valley energies of $E_{2a}$ and $E_{2b}$. 
The width of the valley of $E_{2b}$ is wider than that of $E_{2a}$ because $E_{2b}=E_{2a}+E_{\rm VS}$. 
Then, the shoulder of Fig.~\ref{fignoqubit} is the result of the wide resonant peak added to the narrow resonant peak.

\begin{figure}
\centering
\includegraphics[width=8.5cm]{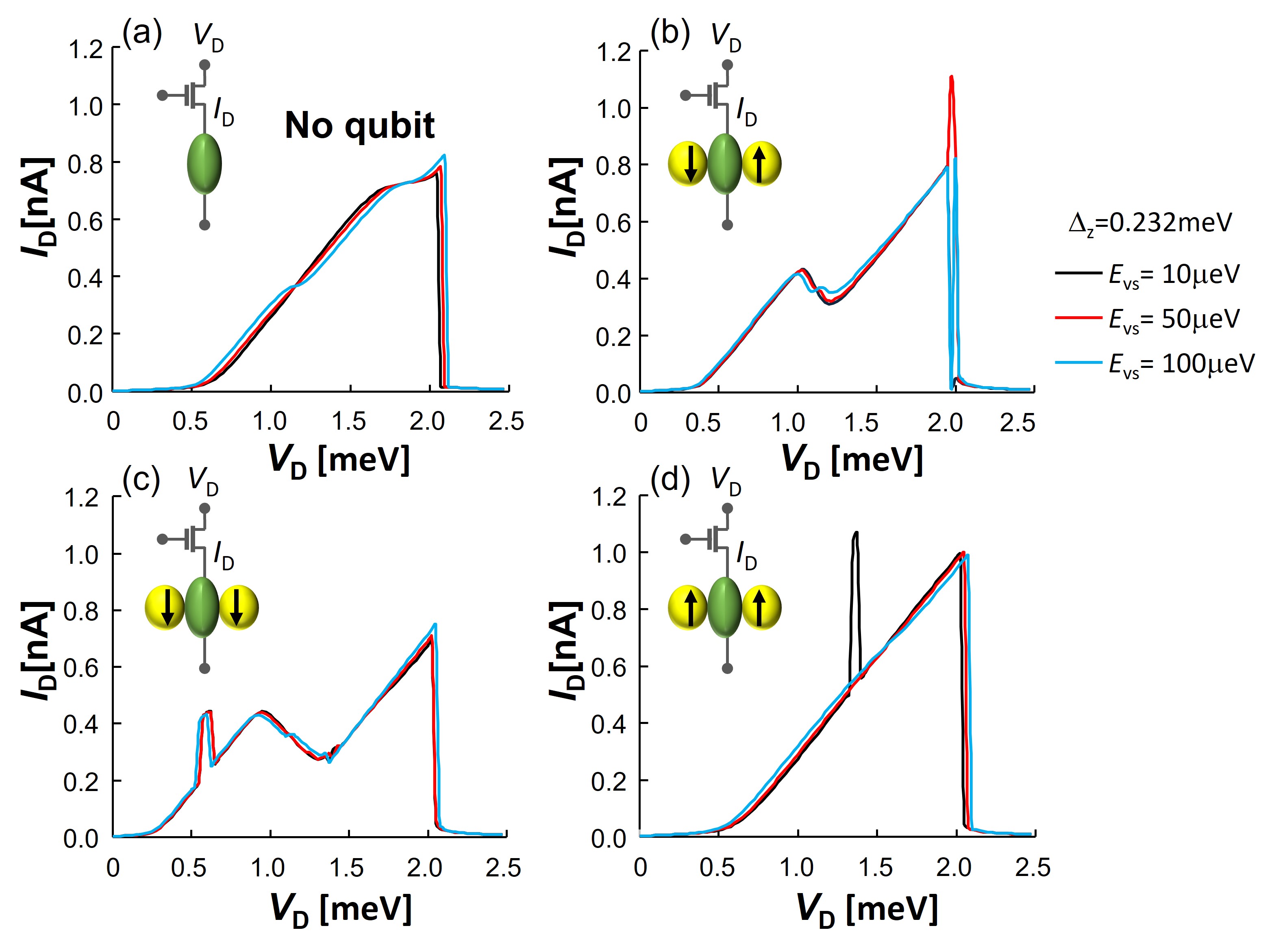}
\caption{
$I_{\rm D}$-$V_{\rm D}$ characteristics in the small valley region ($E_{\rm VS}<\Delta_z$). 
$\Gamma_0=3.0\times 10^{-6}$ eV, $E_F$=1 meV, $T=100$~mK and $E_2^{(0)}=0.88$~meV. 
$E_1=E_3=1.2$ meV, $L=10$~{\textmu}m, and $V_{\rm G}=$0.1~V.
Both QD${}_1$ and QD${}_3$ have the same $E_{\rm VS}$. 
(a) No qubit case, $W=0$. 
(b) Either of QD${}_1$ or QD${}_3$ has a $\upa$-spin state and the other has a $\dna$-spin state. 
(c) Both QD${}_1$ and QD${}_3$ have $\dna$-spin states. 
(d) Both QD${}_1$ and QD${}_3$ have $\upa$-spin states. For (b--d), 
$W=0.09$~meV.
}
\label{figEvs}
\end{figure}
\subsection{Uniform valley splitting}\label{uniform}
Figure~\ref{figEvs} shows $I_{\rm D}$-$V_{\rm D}$ with a connection to the transistor when both sides of the qubit-QDs have the same value of $E_{\rm VS}$. 
The transistor size $L=10$~{\textmu}m is determined to increase the different $V_{\rm out}$ values depending on the qubit states. 
This situation is realized when the resistance of the transistor is comparable to that of the channel-QD. 
Because the current characteristics of the transistor have less nonlinearity than those of the channel-QD, 
the $I_{\rm D}$ values of Fig.~\ref{figEvs} have gentler slopes than those of Fig.~\ref{fignoqubit}(b).

The different $I_{\rm D}$-$V_{\rm D}$ characteristics are explained similarly to Fig.~\ref{figNoTrwithqubit}. 
For the $\dna\dna$ pair ($|00\ra$ state) of Fig.~\ref{figEvs}(c), 
the resonant energy level of the qubit-QDs, 
which is the upper energy level of the singlet state, becomes high, 
as shown in Figs.~\ref{fig3}(a) and \ref{fig3}(c), resulting in a peak structure at lower $V_{\rm D}$. 
Meanwhile, for the $\upa\upa$ pair ($|11\ra$ state) of Fig.~\ref{figEvs}(d), 
the resonant energy level of the qubit-QDs becomes low, as shown in Figs.~\ref{fig3}(b) and \ref{fig3}(d). 
Then, the resonant peak structure is hidden in the original resonant-tunneling 
peak between the source and drain. 
The $I_{\rm D}$-$V_{\rm D}$ characteristics of Fig.~\ref{figEvs}(b) takes the middle properties between Figs.~\ref{figEvs}(c) and \ref{figEvs}(d).
Due to the presence of two energy levels in each QD, multiple resonant peaks appear in the current characteristics. 
The sharp peaks observed in Figs.~\ref{figEvs}(b-d) can be attributed to the complex resonant structure involving six energy levels across the three QDs 
(as indicated in the denominator of Eq.~(\ref{current_formula})). 
Consequently, the presence of valley energy levels enhances the transitions between energy levels, 
resulting in a greater difference in the $I_{\rm D}$s that reflect the qubit states.

\subsection{Nonuniform valley splitting}\label{nonuniform}
\begin{figure}
\centering
\includegraphics[width=8.5cm]{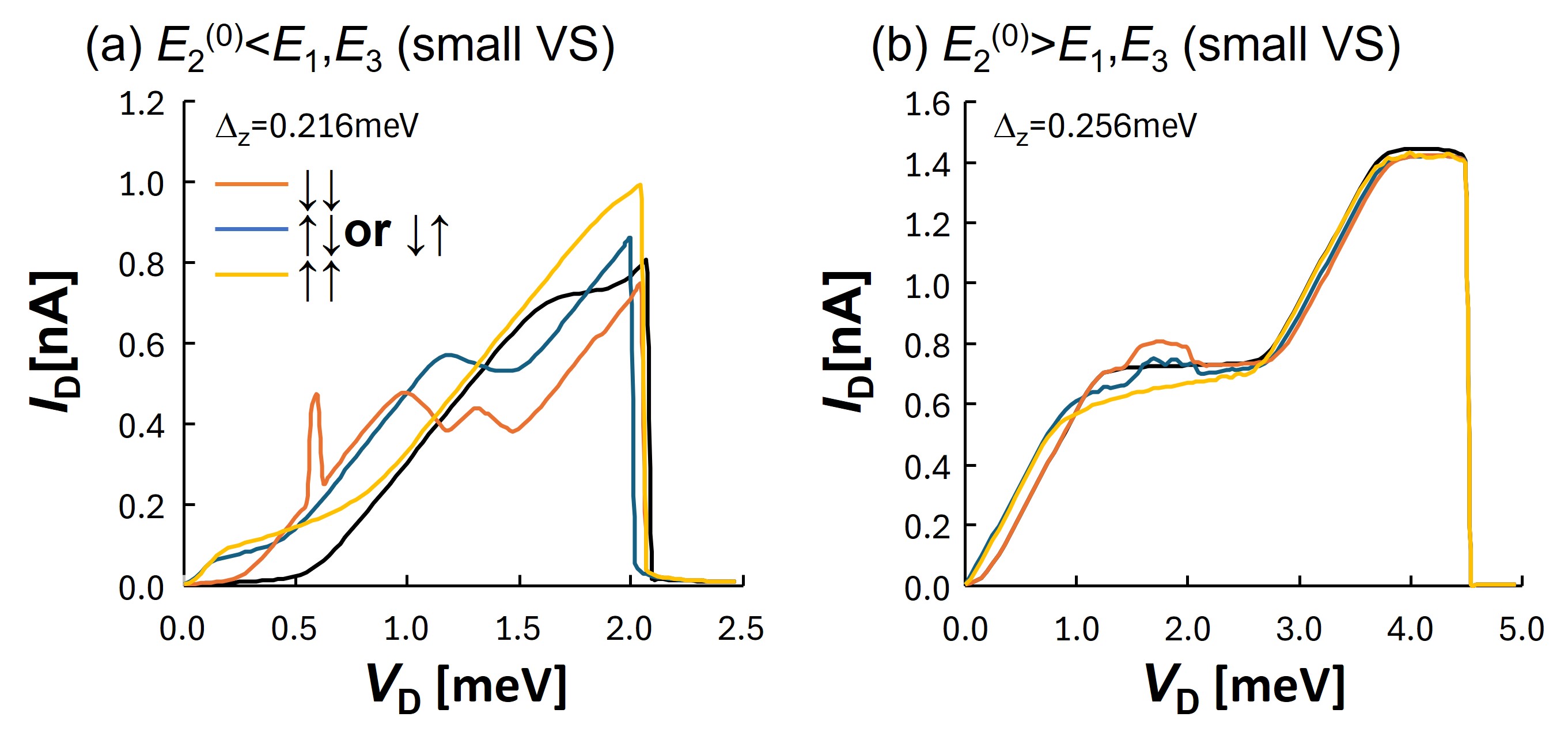}
\caption{
$I_{\rm D}$-$V_{\rm D}$ characteristics in the small valley region ($E_{\rm VS}<\Delta_z$).
$\Gamma_0=3.0\times 10^{-6}$~eV, $E_F$=1~meV, and $T=100$~mK. 
$L=10$~{\textmu}m, $V_{\rm G}=$0.1~V, and $W=0.09$~meV.
$E_{\rm VS1}=10$~{\textmu}eV, $E_{\rm VS2}=20$~{\textmu}eV, and $E_{\rm VS3}=50$~{\textmu}eV.
(a) $E_1=1.2$~meV, $E_2^{(0)}=0.88$~meV, and $E_3=1.4$~meV.
(b) $E_1=1.2$~meV, $E_2^{(0)}=1.4$~meV, and $E_3=1.1$~meV.
}
\label{VSrandomIV}
\end{figure}

It has been reported that valley splitting energies vary depending on the location in the range of $(100{\rm nm})^2$~\cite{Losert}. 
Herein, we consider the case where the three QDs have different valley splitting energies. 
We calculate $E_{\rm VS}=$10, 20, 50, and 100~{\textmu}eV for QD${}_1$(left qubit) and QD${}_3$(right qubit). 
Figure~\ref{VSrandomIV} shows the $I_{\rm D}$-$V_{\rm D}$ characteristics of the small valley region for $W=0.09$~meV.
The shape of the $I_{\rm D}$-$V_{\rm D}$ characteristics is primarily determined by $E_2^{(0)}<E_1, E_3$ (Fig.~\ref{VSrandomIV}(a)) or $E_2^{(0)}>E_1, E_3$ (Fig.~\ref{VSrandomIV}(b)).
Because the energy level of QD${}_2$ increases as $V_{\rm D}$ increases, such as $E_2=E_2^{(0)}+V_{\rm D}/2$, 
the case of $E_2^{(0)}<E_1, E_3$ (Fig.~\ref{VSrandomIV}(a)) shows clear peak structures as a result of energy crossing between $E_2$ and $E_1$, $E_3$. 
Meanwhile, for the case of $E_2^{(0)}>E_1, E_3$, nonlinear characteristics appear as a result of higher-order tunneling, resulting in a vague peak structure in Fig.~\ref{VSrandomIV}(b).
By the nonuniformity of the valley splitting,
$I_{\rm D}$-$V_{\rm D}$ characteristics in Fig.~\ref{VSrandomIV} 
are modified in a more complicated way from those in Fig.~\ref{figEvs}.

\begin{figure}
\centering
\includegraphics[width=8.5cm]{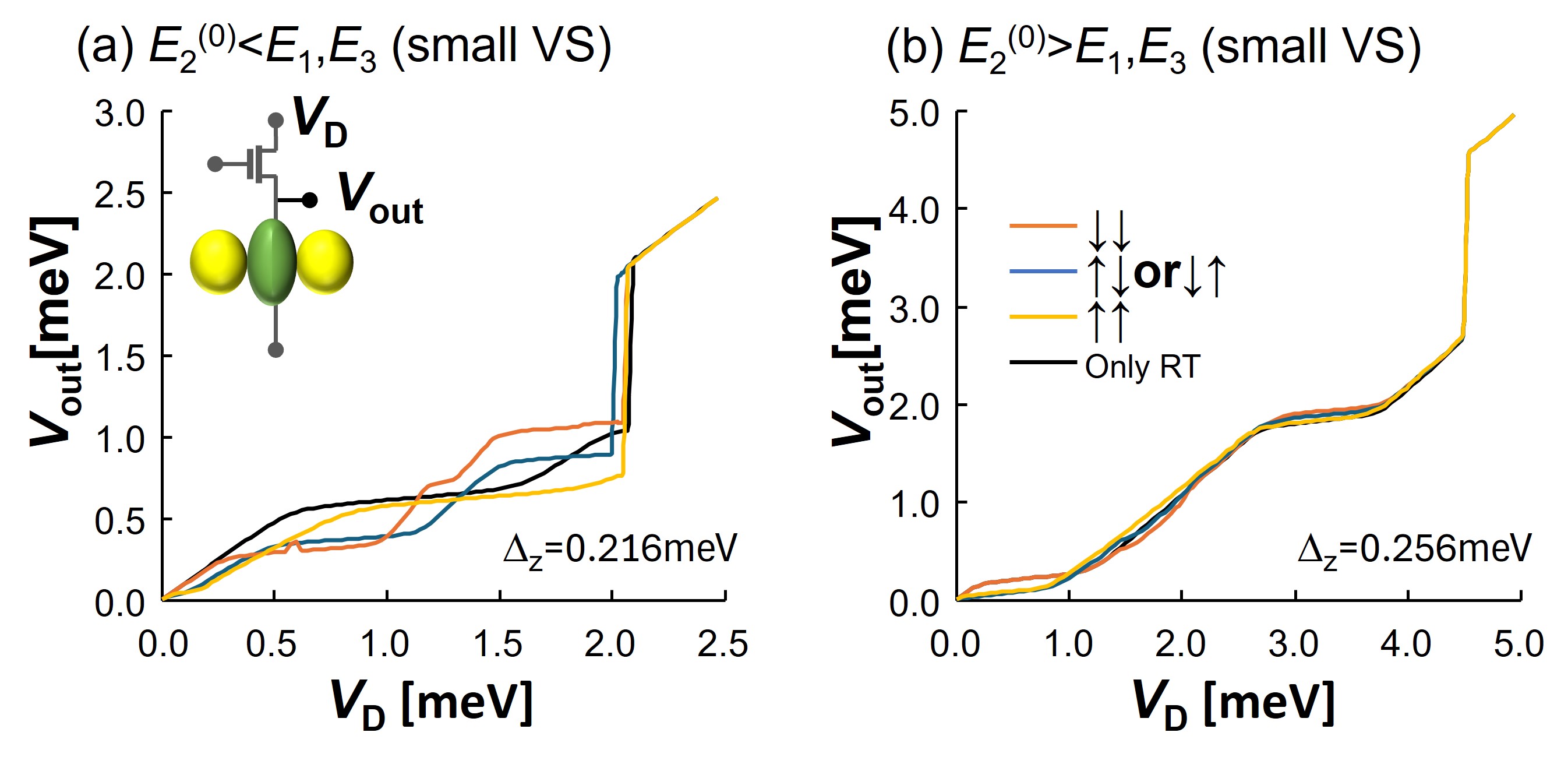}
\caption{
$V_{\rm out}$ as a function of $V_{\rm D}$ in the small valley region.
(a) $E_1=1.2$~meV, $E_2^{(0)}=0.88$~meV, and $E_3=1.4$~meV.
(b) $E_1=1.2$~meV, $E_2^{(0)}=1.4$~meV, and $E_3=1.1$~meV.
The other parameters are the same as those in Fig.~\ref{VSrandomIV}.
}
\label{VSrandomV}
\end{figure}

Figure~\ref{VSrandomV} shows the output voltage $V_{\rm out}$ corresponding to Figs.~\ref{VSrandomIV}(a) and \ref{VSrandomIV}(b). 
For the case of $E_2^{(0)}<E_1, E_3$ (Fig.~\ref{VSrandomV}(a)), a clear separation among various $V_{\rm out}$ values can be observed. 
In contrast, for the case of $E_2^{(0)}>E_1, E_3$ (Fig.~\ref{VSrandomV}(b)), the difference among $V_{\rm out}$ values becomes small. 
A larger voltage difference is desirable for distinguishing different qubit states, which is conducted by connected circuits, as in a previous study~\cite{tanaAPL}. 
Thus, the case of $E_2^{(0)}<E_1, E_3$ is better for the detection of qubit states.

\subsection{Measurement time}\label{time}
\begin{figure}
\centering
\includegraphics[width=8.5cm]{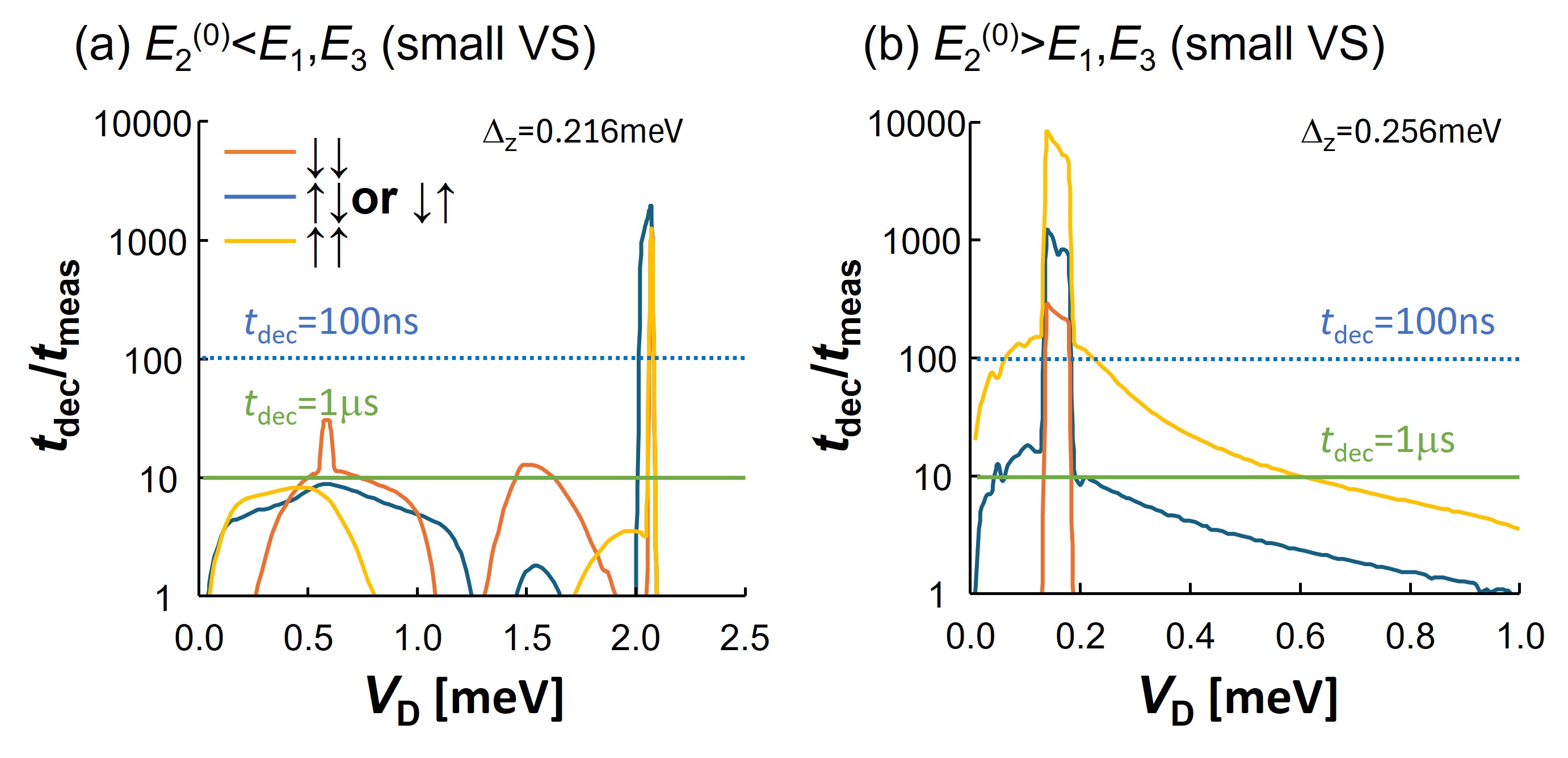}
\caption{
$t_{\rm dec}/t_{\rm meas}$ as a function of $V_{\rm D}$ in the small valley region.
(a) $E_1=1.2$~meV, $E_2^{(0)}=0.88$~meV, and $E_3=1.4$~meV.
(b) $E_1=1.2$~meV, $E_2^{(0)}=1.4$~meV, and $E_3=1.1$~meV.
The other parameters are the same as those in Fig.~\ref{VSrandomIV}.
The dotted horizontal line indicates that more than 100 readouts are possible during $t_{\rm dec}=100$~ns above this line.
The lower solid line indicates that more than 100 readouts are possible during $t_{\rm dec}=1$~{\textmu}s above this line.
}
\label{VSrandomM}
\end{figure}

Figure~\ref{VSrandomM} shows the results for $t_{\rm dec}/t_{\rm meas}$. 
Regarding $t_{\rm dec}/t_{\rm meas}$, simultaneous peaks of $t_{\rm dec}/t_{\rm meas}>100$ appear around the right end of the resonant peak for the case of $E_2^{(0)}<E_1, E_3$. 
In contrast, for the case of $E_2^{(0)}>E_1, E_3$, 
a large $t_{\rm dec}/t_{\rm meas}$ can be seen around $V_{\rm D}=0$, where the energy levels of the three QDs are close to each other. 
The region of $t_{\rm dec}/t_{\rm meas}>100$ for $W=0.18$~meV appears to be smaller than that for $W=0.09$~meV (figure not shown).
Increased mixing of resonant energy levels appears to increase the nonlinearity of $I_{\rm D}$-$V_{\rm D}$ and increase $t_{\rm dec}/t_{\rm meas}$.
The valley splitting energy is closer to $W=0.09$~meV than $W=0.18$~meV,
and increased energy level mixing is expected at $W=0.09$~meV.
Note that $t_{\rm dec}/t_{\rm meas}$ in Fig.~\ref{VSrandomM} is larger 
than that in Fig.~\ref{apfig7} of appendix~\ref{app:large_valley}, which is equivalent  to no valley splitting case.
Therefore, the increased nonlinearity of the current improves the readout 
in this small nonuniformity of the valley energy levels.

We also calculate $E_{\rm VS1}=10$~{\textmu}eV, $E_{\rm VS2}=50$~{\textmu}eV, 
$E_{\rm VS3}=100$~{\textmu}eV, and $E_{\rm VS1}=10$~{\textmu}eV, $E_{\rm VS2}=50$~{\textmu}eV, $E_{\rm VS3}=10$~{\textmu}eV, both for $E_2^{(0)}<E_1, E_3$. 
In the former case, we can see the simultaneous peaks $t_{\rm dec}/t_{\rm meas}>100$ around the right end of the resonant peak around $V_{\rm D}\approx 2.0V$ (figures not shown). 
However, in the latter case, the peaks $t_{\rm dec}/t_{\rm meas}>100$ around $V_{\rm D}\approx 2.0V$ deviate from each other (figures not shown). 
Therefore, although it is difficult to obtain a unified understanding of these cases, 
simultaneous peaks with $t_{\rm dec}/t_{\rm meas}>100$ are expected 
when the valley splitting energies between the nearest QDs are close to each other.


\section{Discussions}~\label{sec:discussions}
In this study, 
we focused on the relationship between the coherence time $t_{\rm dec}$ and the measurement time $t_{\rm meas}$ and did not mention how to determine $|0\ra$ and $|1\ra$ in the succeeding circuit, 
which is connected to the transistor in Fig.~\ref{fig1}. 
Thus, the ratio $t_{\rm dec}/t_{\rm dec}$ does not exactly represent the required number of readouts. 
In most cases, the qubit state could be determined by taking an average of repeated measurements. 
If the transistor in Fig.~\ref{fig1} is connected to the sense amplifier such as those in a previous study~\cite{tanaAPL}, a single-shot readout would be possible. 
For example, $t_{\rm dec}/t_{\rm meas}\approx 100$ with $t_{\rm dec}=100$~ns means $t_{\rm meas}\approx 1$~ns. 
Because the switching speed of a conventional CMOS is in the order of ps~\cite{Jena}, the resolution of the output signal by CMOS is higher than the changing time of qubit states, 
and we can detect the change in the $V_{\rm out}$ using the CMOS circuits. 
Further analysis regarding the effect of noise by transistors is the future issue.

The electron that enters into the qubit-QD forms a singlet state with the electron in the qubit. 
If there is no energy relaxation, the electron with the same spin as that when it enters the qubit-QD goes out to the channel-QD because of energy conservation. 
However, when there is a relaxation process, the qubit state changes. 
In this study, this process was represented by the static decoherence time of $t_{\rm dec}$. 
The dynamic description of this process is beyond the scope of this study and is a problem for future studies.
 
The numerical results of the large valley splitting with the nonuniformity in energy levels are presented in the appendix~\ref{app:large_valley}. 
$t_{\rm dec}/t_{\rm meas}$ becomes smaller with the existence of nonuniformity. 
This indicates that the nonuniform energy levels degrade the performance of the readout process. 
However, because of the nonlinear effects of the multiple energy levels, 
the reduction in $t_{\rm dec}/t_{\rm meas}$ by the valley splitting seems to be small as long as the nonuniformity is small. 
The current characteristics under the nonuniformity strongly depend on the system parameters (e.g., $E_{\rm VS}$, $W_{ij}$, $E_i$, and $\Gamma_i$). 
Thus, the maximum tolerance to the nonuniformity should be determined depending on each system. 
Hence, systematic estimation is also an issue for future research.

\section{Conclusions}~\label{sec:conclusions}
In summary, we theoretically investigated the effect of valley splitting on the readout process of spin qubits constructed in a three-QD system, 
aiming at future two-dimensional qubit arrays. 
The current formula (Eq.(\ref{current_formula})) was derived on the basis of the nonequilibrium Green function method by including the mixing of the energy levels in the QDs. 
In the small valley splitting region ($E_{\rm VS}<\Delta_z$) and large valley splitting region ($E_{\rm VS}>\Delta_z$), 
we found the parameter region where more than a hundred readouts are expected ($t_{\rm dec}/t_{\rm meas}>100$). 
In particular, increased mixing of the resonant energy levels enhances the nonlinearity of the current \(I_{\rm D}\) and improves the readout in the small valley region (larger \(t_{\rm dec}/t_{\rm meas}\)). The degree of mixing is highly dependent on the device parameters, 
so the optimal values must be determined for each device. 
These results demonstrate that the proposed system can effectively detect the resonant energy levels of the coupled QD system. 
Therefore, while a larger valley splitting energy is desirable for qubit operations, the readout of spin states can still be effectively accomplished, even in regions with valley splitting.
\subsection*{DATA AVAILABILITY}
The data supporting the findings of this study are available in this article.

\begin{acknowledgments}
We are grateful to T. Mori, H. Fuketa, and  S. Takagi for their insightful discussions.
This study was partially supported by the MEXT Quantum Leap Flagship Program (MEXT Q-LEAP), Grant Number JPMXS0118069228, Japan.
Furthermore, this study was supported by JSPS KAKENHI Grant Number JP22K03497.
\end{acknowledgments}

\subsection*{AUTHOR CONTRIBUTIONS}
T.T. conceived and designed the theoretical calculations. 
K.O. discussed the results from the experimentalist viewpoint. 

\subsection*{CONFLICT OF INTEREST}
The authors have no conflicts of interest to disclose.

\appendix
\section{Qubit array aiming at a 3D stacked qubit system}\label{app:array}
Surface code is a critical area of focus for qubit systems, requiring an error rate of less than 1\%~\cite{Fowler}. 
Various proposals for implementing surface codes using spin qubits have been made~\cite{Hill,Veldhorst1,Veldhorst,Li}. 
In Fig.~1(a), it is assumed that independent electrodes are attached to the control gates $V_{\text{cg}, i}$ one by one. 
The wiring for $V_{\text{cg}, i}$s are constructed in parallel to the wiring of the sources and drains.

Our strategy involves leveraging existing commercial transistor architectures to minimize fabrication costs, 
as advanced transistors can be prohibitively expensive. 
By stacking the one-dimensional array shown in Fig.1, 
we can achieve a two-dimensional qubit array suitable for the surface code structure using advanced semiconductor technologies.

\begin{figure}
\centering
\includegraphics[width=8.5cm]{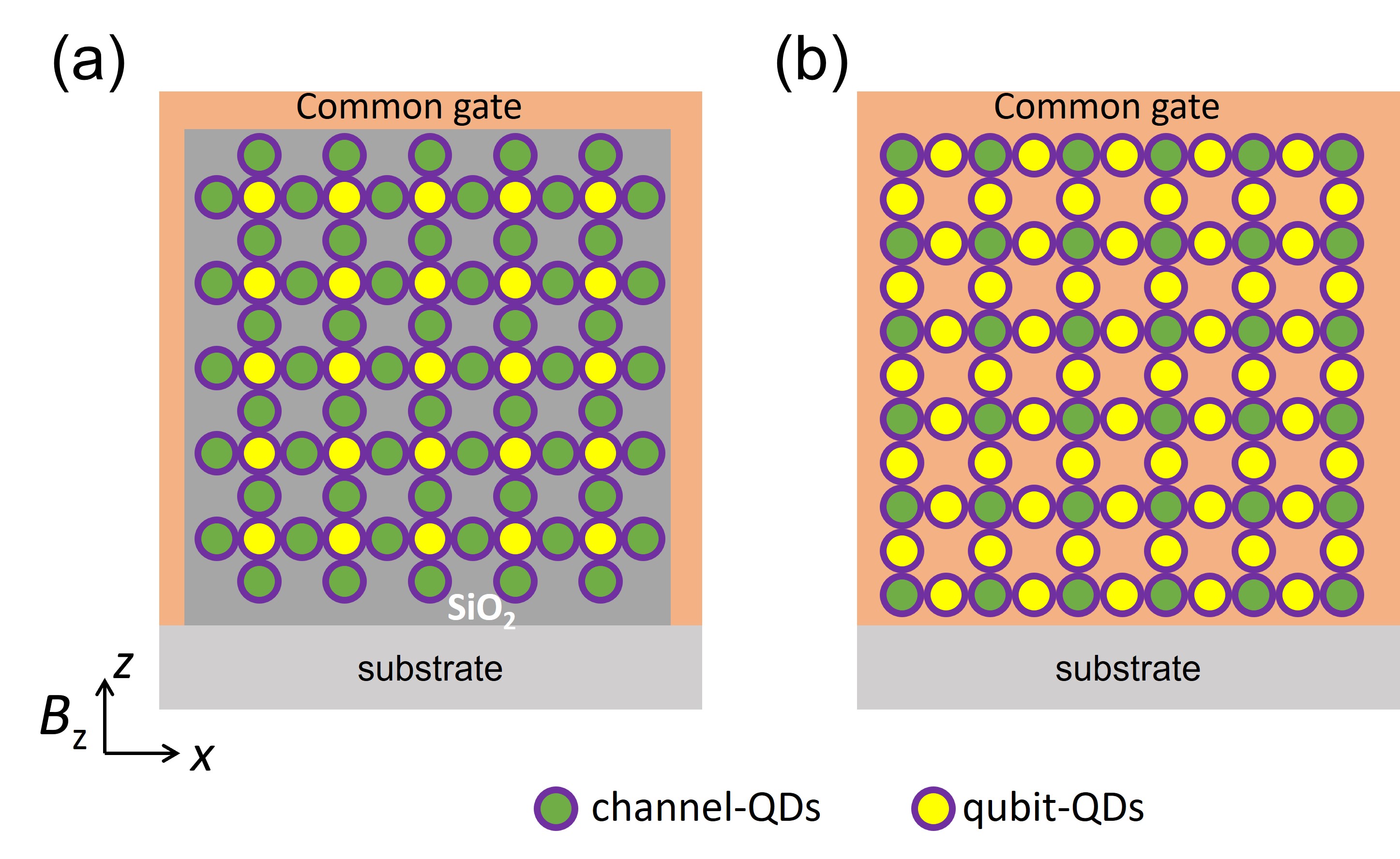}
\caption{
Two types of the stacking structures using the qubit array in Fig.1(a).
Cross section at the center of QDs.
(a) Each qubit is surround by four channel-QDs.
(b) Each channel-QD is surrounded by four qubits.
The yellow circles represent the qubit-QDs, whereas the green ellipse represents the channel-QDs,
which serve as the coupling and readout of the qubits. 
All QDs are surrounded by insulators such as SiO${}_2$.
Spin directions are changed using a magnetic field gradient method, utilizing micromagnets placed above the structure (not shown in the figure).
The static magnetic field $B_z$ is applied along the $z$-direction.
}
\label{figarray1}
\end{figure}

\begin{figure}
\centering
\includegraphics[width=7cm]{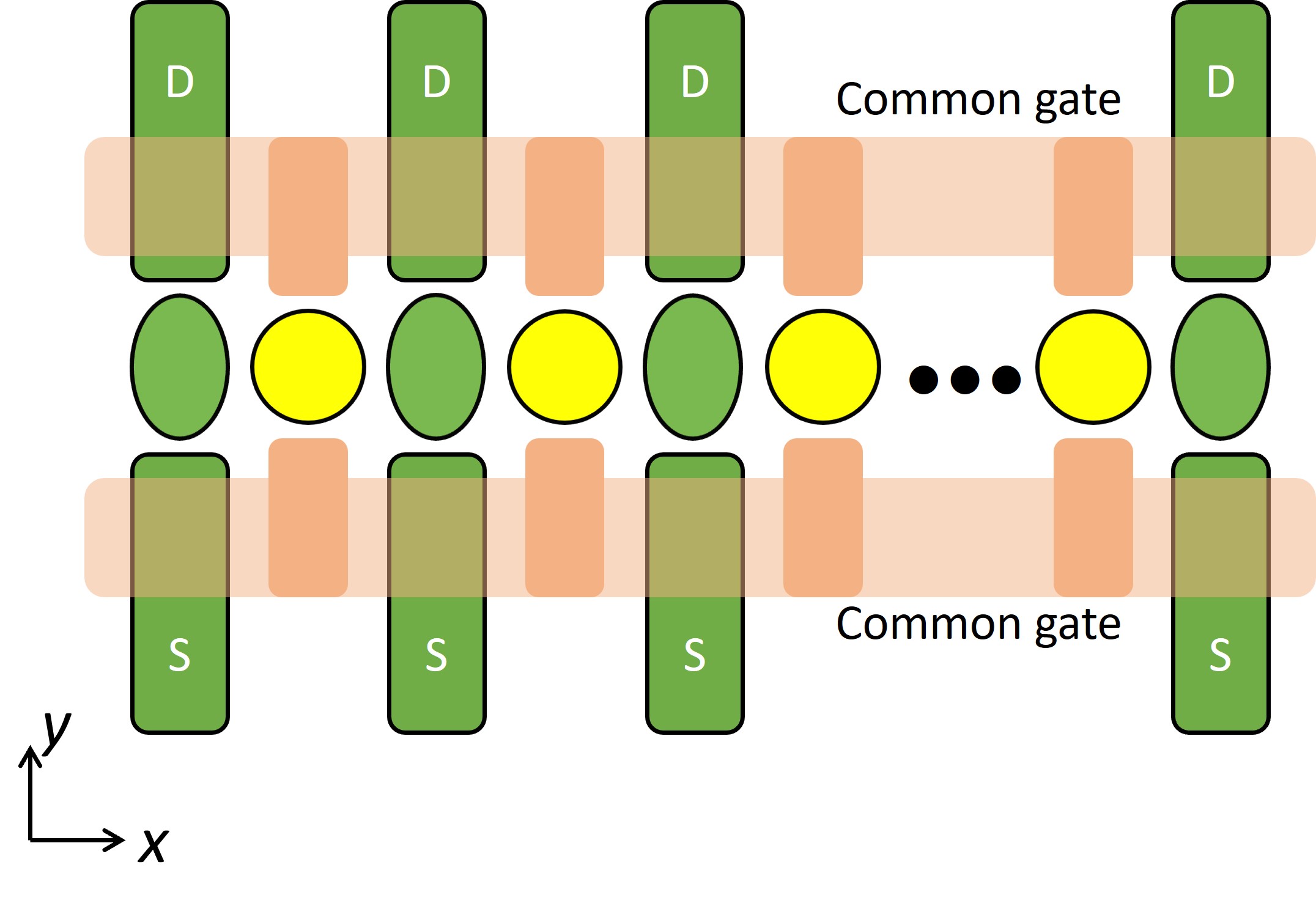}
\caption{
Example of the common gate structure in Fig.~\ref{figarray1}(a).
The common gate is connected to the qubit-QD through the gap between the source-drain wires.
The source-drain wires are also placed over the qubit-QD.
All structures are surrounded by SiO${}_2$.
}
\label{figarray2}
\end{figure}

Here, we consider the case in which the control gates $V_{\text{cg}}$ in Fig. 1(a) are represented by a single common gate. 
Two types of stacking structures are shown in Fig.~\ref{figarray1}. 
In Fig.~\ref{figarray1}(a), each qubit-QD is surrounded by four channel-QDs, as proposed in Ref.~\cite{tanaJAP}. 
In this configuration, each qubit can be controlled by the four channel-QDs, resulting in high controllability. 
However, the area that the common gate covers over each qubit QD is small, as shown in Fig.~\ref{figarray2}, 
meaning that gate controllability will depend on the detailed 3D structure.

In Fig.~\ref{figarray1}(b), qubits are arranged diagonally, resulting in coupling between diagonally positioned qubits. 
In this case, the four qubits surrounding each channel QD must have different energy levels 
to independently control qubit-qubit coupling and individual qubit. 
On the other hand, the electric potential of the qubits can be effectively controlled 
because each qubit is surrounded by four directions of influence from the common gate.
Currently, the optimal structure for realizing the design in Fig. 1 is a future issue. 
To further refine our design, we have initiated technical CAD simulations, as detailed in Ref.~\cite{tanaTCAD}.

\begin{figure}
\centering
\includegraphics[width=7cm]{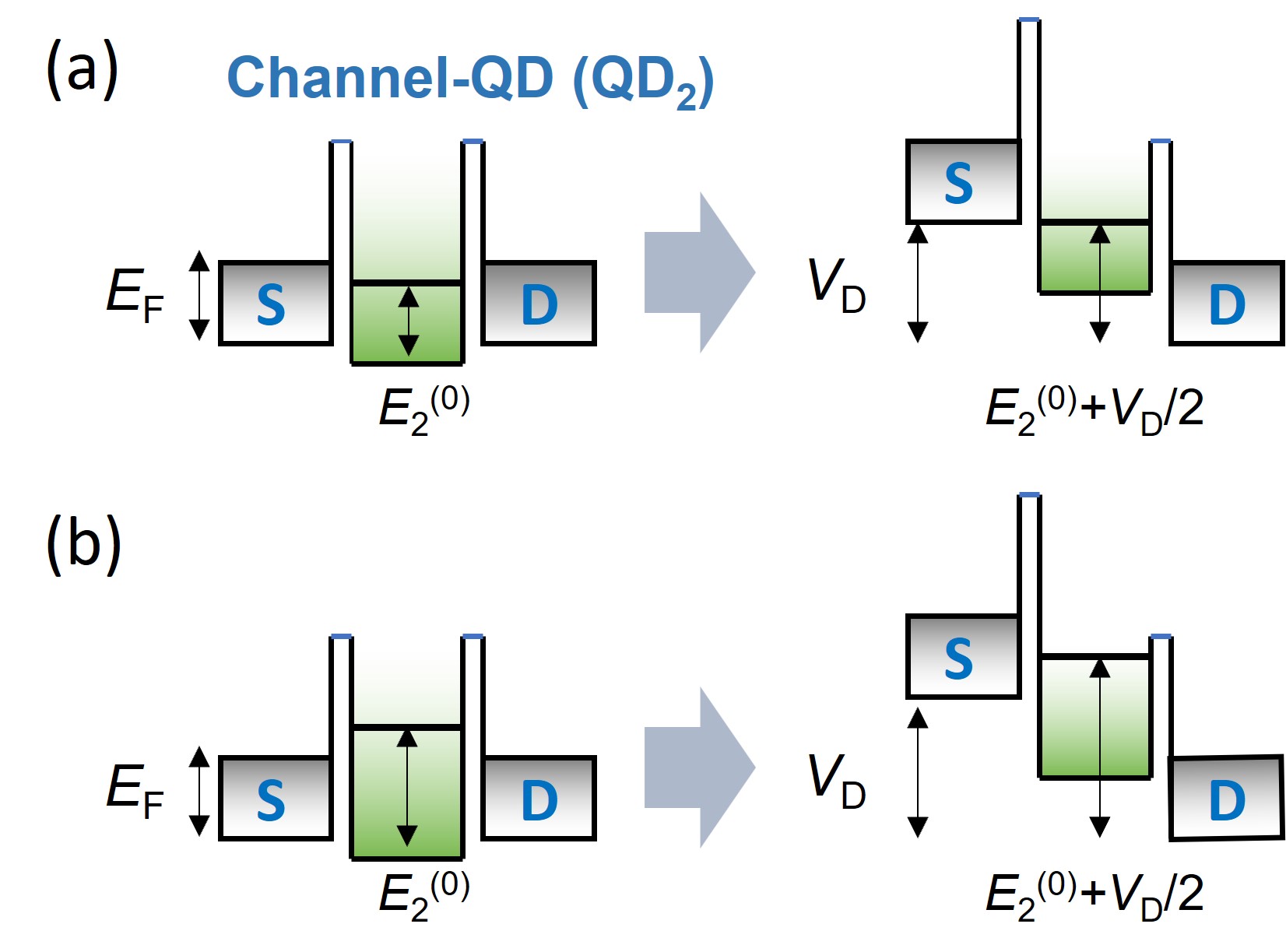}
\caption{
Simple analysis of the resonant-tunneling phenomena.
Two situations are observed depending on the relative energy between the electrode 
and channel-QD (QD${}_2$). 
(a)$E_2^{(0)}>E_F$. (b) $E_2^{(0)}<E_F$.
The result is listed in Table~\ref{tbl1}.
}
\label{appresonant}
\end{figure}

\section{Analysis of resonant-tunneling without qubit}\label{analysis_of_rt}
\subsection{No magnetic field case}
In this section, we will analyze the simple resonant peak structure without qubits or a magnetic field.
The general trend of the peak current can be understood by examining the simple band structure.
Two different cases are considered, based on whether $E_2^{(0)}<E_F$ or $E_2^{(0)}>E_F$, as shown in Fig.~\ref{appresonant}.
When the resonant current begins to flow, the energy level of the source surpasses that of the channel-QD (QD$_2$). 
This can be represented as follows:
\begin{equation}
E_F+V_{\rm D} > E_2^{(0)}+V_{\rm D}/2. 
\label{analysis1}
\end{equation}
Moreover, the energy of the channel-QD should exceed the drain energy level, resulting in:
\begin{equation}
E_2^{(0)}+V_{\rm D}/2 >E_F. 
\label{analysis2}
\end{equation}
These result in the following:
\begin{equation}
V_{\rm D} > 2|E_F-E_2^{(0)}|. 
\label{analysis3}
\end{equation}

The resonant peak disappears when $E_2^{(0)}+V_{\rm D}/2$ 
is less than that at the bottom of the source energy level(right side of Fig.~\ref{appresonant}(a)), which is expressed as follows:
\begin{equation}
E_2^{(0)}+V_{\rm D}/2 <V_{\rm D}.
\label{rt2}
\end{equation}
Therefore, the bias voltage $V_{\rm D}^{\rm off}$ representing the disappearance of the resonant peak is expressed as follows:
\begin{equation}
V_{\rm D}^{\rm off}= 2 E_2^{(0)}.
\label{rt3}
\end{equation}

Thus, for $E_2^{(0)}<E_F$(Fig.~\ref{appresonant}(a)), the resonant peak region is approximately expressed as follows:
\begin{equation}
2(E_F-E_2^{(0)})< V_{\rm D}< 2E_2^{(0)}, 
\end{equation}
where $V_{\rm D}^{\rm on}$ is expressed as follows: 
\begin{equation}
V_{\rm D}^{\rm on} =2(E_F-E_2^{(0)}). \label{rt1}
\end{equation}
The approximate values of the width and center of the resonant peak are 
$4E_2^{(0)}-2E_F$ and $E_F$, respectively.

For $E_2^{(0)}>E_F$ (Fig.~\ref{appresonant}(b)), we have
\begin{equation}
2(E_2^{(0)}-E_F)< V_{\rm D}< 2E_2^{(0)}.
\end{equation}
where $V_{\rm D}^{\rm on}$ is expressed as follows: 
\begin{equation}
V_{\rm D}^{\rm on} =2(E_2^{(0)}-E_F). 
\end{equation}
The approximate values of the width and center of the resonant peak are 
$E_F$ and $4E_2^{(0)}-2E_F$, respectively.
The results are listed in Table I.

\begin{table}[htbp]
\caption{Analysis of resonant-tunneling structure without magnetic field.}
\begin{center}
	\begin{tabular}{c|c|c|c}
		\hline
		Situations & Peak region & Peak width & Peak center \\
		\hline
		$E_2^{(0)}\!<\!E_F$ & $2(E_F\!-\!E_2^{(0)})\!<\! V_{\rm D}\!<\! 2E_2^{(0)}$ & $4E_2^{(0)}\!\!-\!2E_F$ 
		& $E_F$\\
		$E_2^{(0)}\!>\!E_F$ & $2(E_2^{(0)}\!\!-\! E_F)\!<\! V_{\rm D}\!<\! 2E_2^{(0)}$ & $2E_F$ 
		& $2E_2^{(0)}\!\!-\! E_F$ \\
		\hline
	\end{tabular}
	\label{tbl1}
	\end{center}
\end{table}

\subsection{Finite magnetic field case}
When a magnetic field is present, the $\upa$ and $\dna$ spin currents can be analyzed by adjusting the Fermi energy to $E_F\mp \Delta_z/2$, 
as discussed in the main text.
Here, we will focus on the $\upa$-current case (the $\dna$-current case can be addressed by substituting $\Delta_z$ by $-\Delta_z$). 
The resonant-tunneling $\upa$-current can be approximated as outlined in Table~\ref{tbl2}, where $E_F$ in Table~\ref{tbl1} is replaced by $E_F- \Delta_z/2$.
For example, when $E_2^{(0)}+\Delta_z/2 < E_F$ is satisfied, 
the bias voltage at which the resonant peak appears is expressed as follows:
\begin{equation}
V_{\rm D\upa}^{\rm on} =2(E_F-E_2^{(0)}-\Delta_z/2). 
\label{rt4}
\end{equation}
The end of the resonant peak is expressed as follows:
\begin{equation}
V_{{\rm D}\upa}^{\rm off}= 2 E_2^{(0)}.
\label{rt5}
\end{equation}
Correspondingly, the resonant peak region is expressed as:
\begin{equation}
2(E_F-E_2^{(0)}-\Delta_z/2)< V_{\rm D}< 2E_2^{(0)}.
\end{equation}
The width and center of the resonant peak are approximated as  $4E_2^{(0)}-2E_F+\Delta_z$ and $E_F-\Delta_z/2$, respectively.

\begin{table*}[htbp]
\caption{Analysis of the resonant-tunneling structure of the $\upa$-current with magnetic field.
The results of the $\dna$-current are obtained by replacing $\Delta_z$ by $-\Delta_z$.}
\begin{center}
 \begin{tabular}{c|c|c|c}
  \hline
  Situations & Peak region & Peak width & Peak center \\
  \hline
  $E_2^{(0)}+\Delta_z/2\!<\! E_F$ & $2(E_F\!-\! E_2^{(0)}\!-\Delta_z/2)\!<\! V_{\rm D}\!<\! 2E_2^{(0)}$ 
  & $4E_2^{(0)}\!\!+\Delta_z-\!2E_F$ & $E_F-\Delta_z/2$\\
  $E_2^{(0)}+\Delta_z/2\!>\! E_F$ & $2(E_2^{(0)}+\Delta_z/2\!\!-\! E_F)\!<\! V_{\rm D}\!<\! 2E_2^{(0)}$ 
  & $2E_F-\Delta_z $ & $2E_2^{(0)}\!\!+\Delta_z/2-\!E_F$ \\
 \hline
 \end{tabular}
 \label{tbl2}
 \end{center}
\end{table*}

\section{Nonuniform large valley region}\label{app:large_valley}
\begin{figure}
\centering
\includegraphics[width=8.5cm]{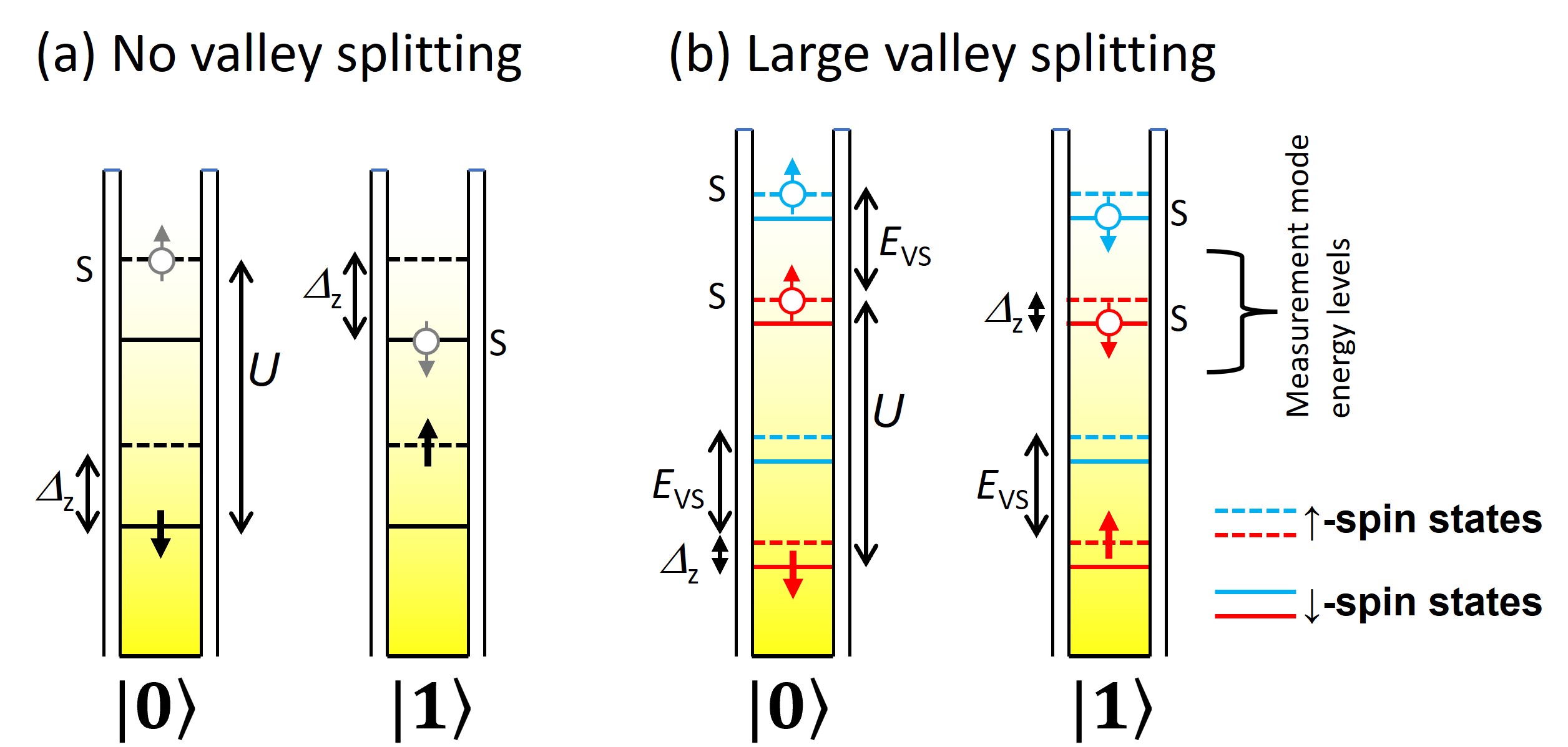}
\caption{Energy band diagram of the two qubit states ($|0\ra$ or $|1\ra$). 
(a) No valley splitting case. (b) Large valley case. 
The solid arrows denote the spins of the qubits (first electron). 
The arrows with white circles represent the possible spin state in which the second electron 
enters the QDs. 
In the large valley region, the tunneling possibilities are similar to those of the no valley splitting case.
}
\label{apfig2}
\end{figure}

\begin{figure}
\centering
\includegraphics[width=8.5cm]{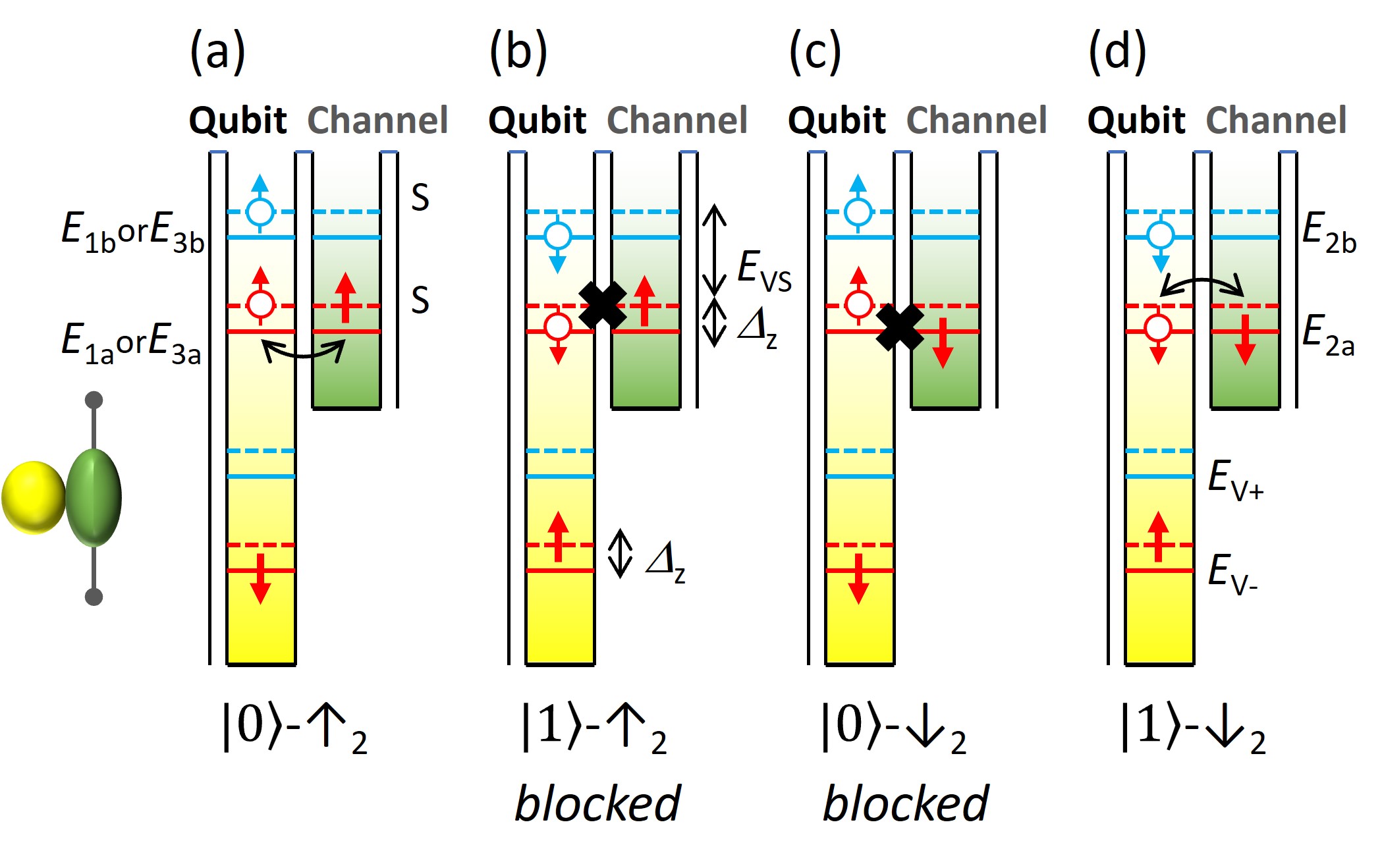}
\caption{Tunneling profile between the channel-QD(QD${}_2$) and qubit-QD for large valley splitting.
The tunneling possibilities are the same as those in small valley splitting (Fig.~\ref{fig3}).
Here, the qubit-QD represents both QD${}_1$ and QD${}_3$.
(a) and (b) represent the $\upa$-current. (c) and (d) represent $\dna$-current.
From (a) and (b), $\upa$-current can exchange $\upa$-spin electron only when the qubit state is $|0\ra$.
From (c) and (d), $\dna$-current can exchange $\dna$-spin electron only when the qubit state is $|1\ra$. 
}
\label{apfig3}
\end{figure}

In this section, we explore the region of nonuniform large valley splitting ($E_1\neq E_3$ and $E_{\rm VS}>\Delta_z$).
Owing to valley splitting,
the energy diagram of qubit-QDs changes from 
Fig.~\ref{apfig2}(a) to Fig.~\ref{apfig2}(b).
Consequently, in cases of large valley splitting, 
the tunneling restrictions remain unchanged compared with no valley splitting case (Fig.~\ref{apfig3}).

\begin{figure}
\centering
\includegraphics[width=8.5cm]{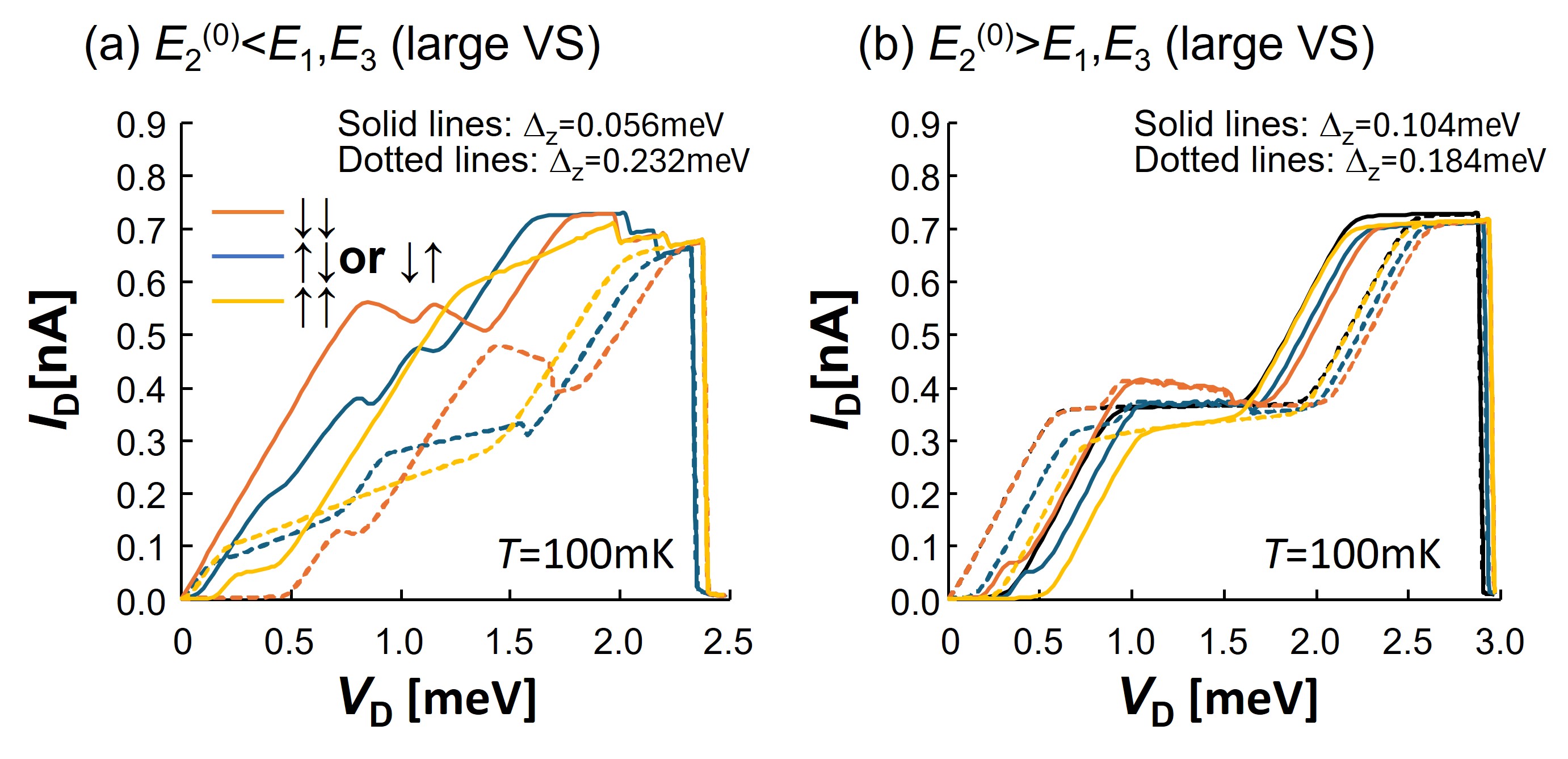}
\caption{$I_{\rm D}$-$V_{\rm D}$ characteristics in the small valley region.
$\Gamma_0=3.0\times 10^{-6}$ eV, $E_F$=1 meV, $T=100$mK.
$L=10\mu$ m, $W=0.18$ meV, and $V_G=$0.1 V. 
(a) $E_1=1.2$ meV, $E_2^{(0)}=1.1$ meV, and $E_3=1.4$ meV.
(b) $E_1=1.2$ meV, $E_2^{(0)}=1.4$ meV, and $E_3=1.1$meV.
}
\label{apfig5}
\end{figure}

\begin{figure}
\centering
\includegraphics[width=8.5cm]{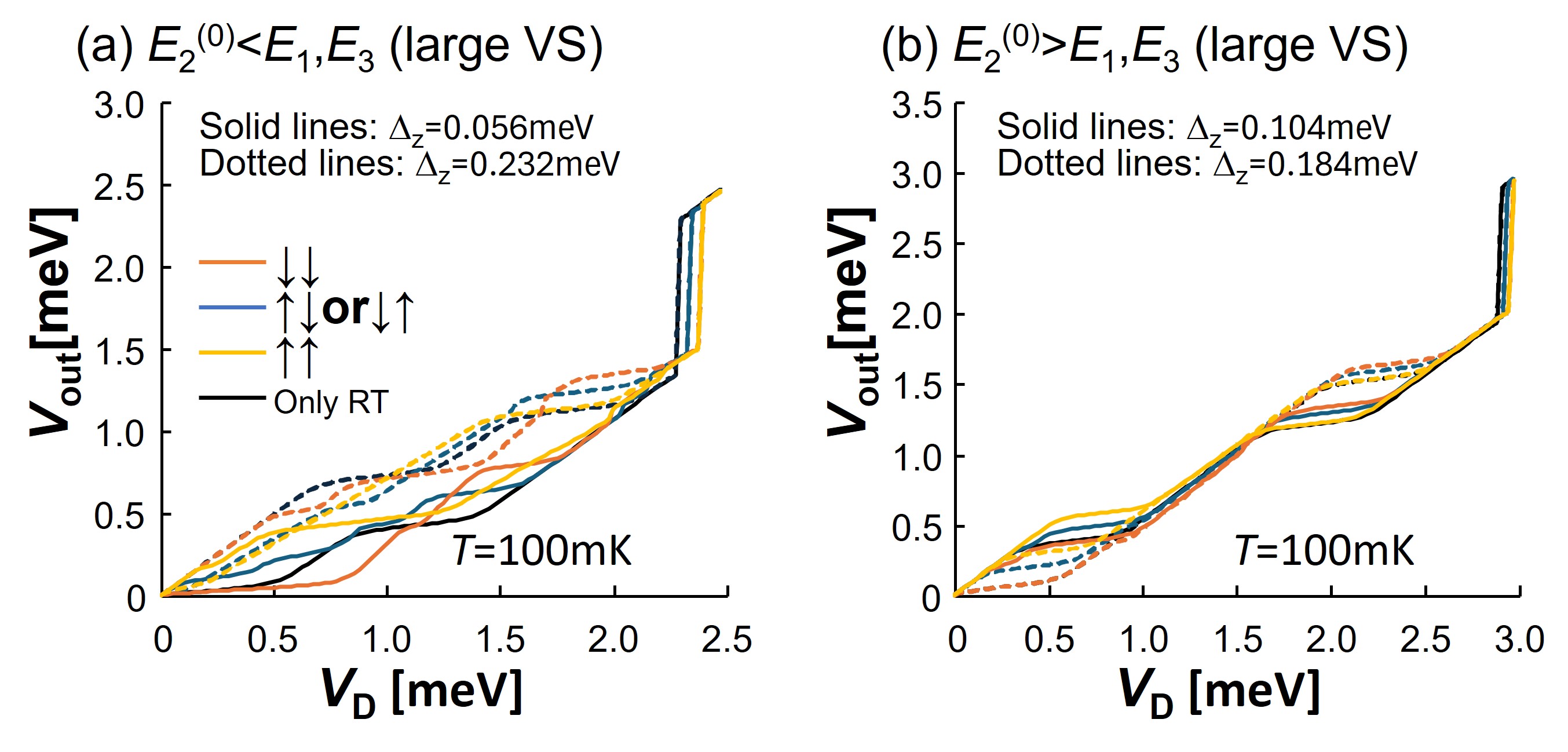}
\caption{
$V_{\rm out}$ as a function of $V_{\rm D}$ in the large valley region.
(a) $E_1=1.2$ meV, $E_2^{(0)}=1.1$ meV, and $E_3=1.4$ meV.
(b) $E_1=1.2$ meV, $E_2^{(0)}=1.4$ meV, and $E_3=1.1$ meV.
Other parameters align with those in Fig.~\ref{apfig5}.
}
\label{apfig6}
\end{figure}

\begin{figure}
\centering
\includegraphics[width=8.5cm]{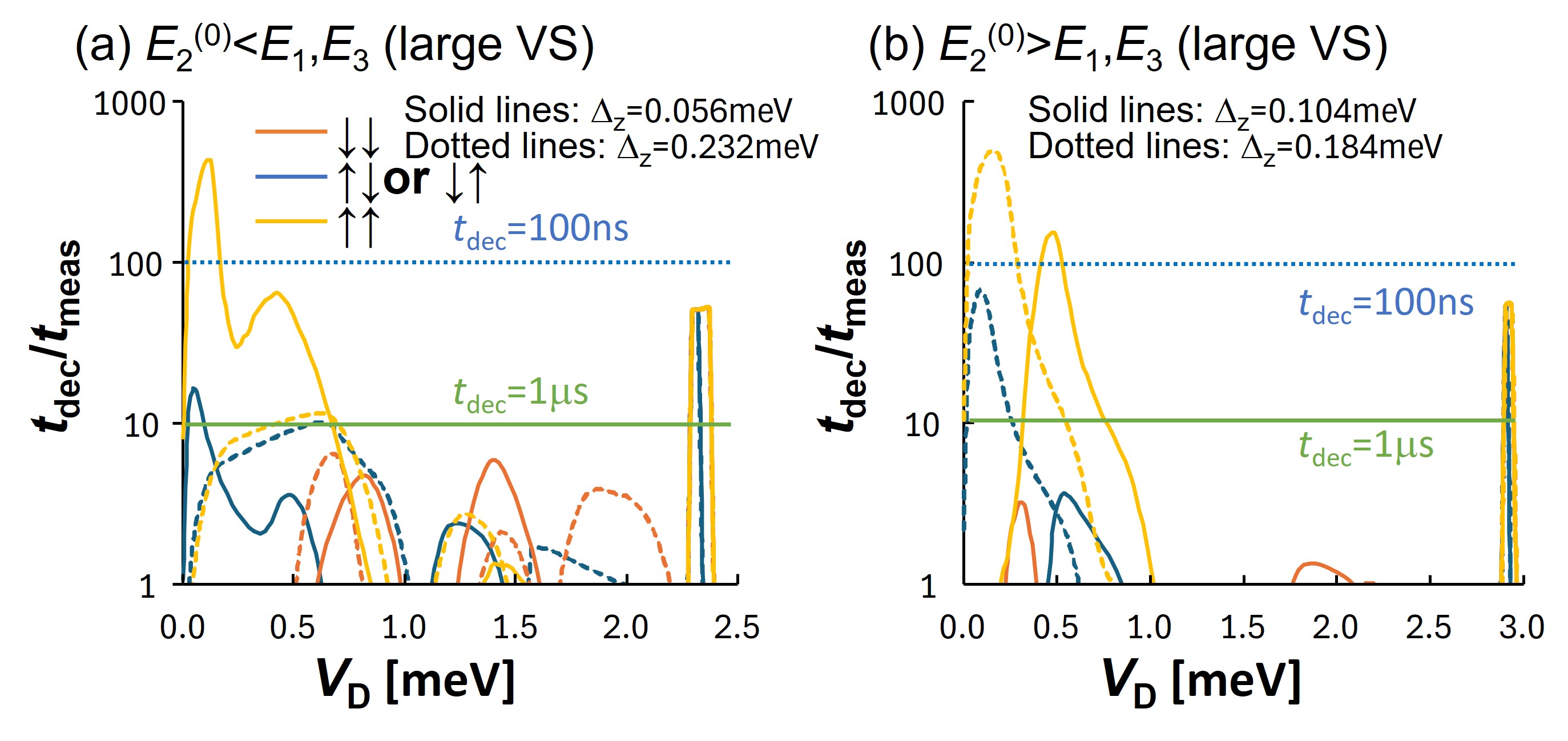}
\caption{
$t_{\rm dec}/t_{\rm meas}$ as a function of $V_{\rm D}$ in the large valley region.
(a) $E_1=1.2$ meV, $E_2^{(0)}=1.1$ meV, and $E_3=1.4$ meV.
(b) $E_1=1.2$ meV, $E_2^{(0)}=1.4$ meV, and $E_3=1.1$ meV.
Other parameters align with those in Fig.~\ref{apfig5}.
The dotted horizontal line indicates that more than 100 readouts are possible during the $t_{\rm dec}=100$ ns above this line.
The lower solid line indicates that over 100 readouts are possible during the $t_{\rm dec}=1\mu$s above this line.
}
\label{apfig7}
\end{figure}

The $I_{\rm D}$-$V_{\rm D}$, $V_{\rm out}$, and $t_{\rm dec}/t_{\rm meas}$ of the $E_{\rm VS}>\Delta_z$ and $E_{\rm VS}>V_{\rm D}$ are shown in Figs.~\ref{apfig5}-\ref{apfig7}.
The difference between Figs.~\ref{apfig5} (a) and (b) lies in the condition of $E_2^{(0)}<E_1,E_3$ (Fig.~\ref{apfig5}(a)) and $E_2^{(0)}>E_1,E_3$ (Fig.~\ref{apfig5}(b)).
As the energy level of QD${}_2$ increases with $V_{\rm D}$, as indicated by $E_2=E_2^{(0)}+V_{\rm D}/2$,
a clear resonant peak is observed for $E_2^{(0)}<E_1,E_3$ (Fig.~\ref{apfig5}(a)).
The resonant peak structure is influenced by the spin states of QD${}_1$ and QD${}_3$.
Therefore, depending on the qubit state (Fig.~\ref{apfig3}), different $I_{\rm D}$s are obtained.
This difference in current results in different measurement times. 
For $E_2^{(0)}>E_1,E_3$(Fig.\ref{apfig5}(b)), 
the variation in $I_{\rm D}$–$V_{\rm D}$ characteristics among different qubit states is not pronounced as when $E_2^{(0)}<E_1,E_3$(Fig.\ref{apfig5}(a)).
However, upon comparing Fig.\ref{apfig5}(b) with Fig.\ref{VSrandomIV}(b), the differences in $I_{\rm D}$-$V_{\rm D}$ characteristics increases slightly 
for the large valleys.

The $V_{\rm out}$ for a gate length $L=10~\mu$m of the transistor is shown in Fig.~\ref{apfig6}. 
Notably, $E_2^{(0)}<E_1,E_3$ (Fig.~\ref{apfig6}(a)) is superior to 
$E_2^{(0)}>E_1,E_3$ (Fig.~\ref{apfig6}(b)) for the detection.

Figure~\ref{apfig7} shows the results for $t_{\rm dec}/t_{\rm meas}$ for the nonuniform large valley splittings. 
Above the blue dotted line, $t_{\rm dec}/t_{\rm meas}$ exceeds 100 for $t_{\rm dec}=100$ ns.
Above the solid horizontal line, $t_{\rm dec}/t_{\rm meas}$ exceeds 100 for $t_{\rm dec}=1 \mu$s. 
For the uniform valley splitting energies ($E_1=E_3=1.2$ meV and $E_2^{(0)}=1.1$ meV), 
the region of $t_{\rm dec}/t_{\rm meas}>100$ for $t_{\rm dec}=100$ ns 
to all qubit states is found near $V_{\rm D}\approx 2.3$ meV
(figures not shown).
However, in Fig.~\ref{apfig7}, there is no region 
where $t_{\rm dec}/t_{\rm meas}>100$ for $t_{\rm dec}=100$ ns
is satisfied  for all qubit states.
Therefore, the nonuniform energy levels degrade the $t_{\rm dec}/t_{\rm meas}$.
If we can increase the coherence time to $t_{\rm dec}=1\mu$s, 
we can find the region of $t_{\rm dec}/t_{\rm meas}>100$ for all qubit states in Figs.~\ref{apfig7}(a) and (b).



\section{Derivation of Green functions}\label{app:green_function}
The equation of motion method can be utilized to derive various Green functions. 
In the following theoretical approach using nonequilibrium Green functions, 
the electrodes are referred to as the left and right electrodes, 
denoted as $L$ and $R$ instead of $S$ and $D$, respectively. 

The equation for the operator $d_{1a}(\omega)$ is given by
\begin{eqnarray}
\lefteqn{
\omega d_{1a}(\omega) = [d_{1a}(\omega),H ] }\nnm\\
&=& E_{1a}d_{1a} 
+ W_{12aa} d_{2a} +W_{12ab} d_{2b}.
\label{d1aw}
\end{eqnarray}
Similarly, we have
\begin{eqnarray}
(\omega -E_{1b})d_{1b} &=& W_{12bb} d_{2b}+W_{12ba} d_{2a}, \nnm \\
(\omega -E_{3a})d_{3a} &=& W_{32aa} d_{2a}+W_{32ab} d_{2b},  \nnm \\
(\omega -E_{3b})d_{3b} &=& W_{32bb} d_{2b}+W_{32ba} d_{2a},  \nnm \\
(\omega -E_{2a})d_{2a} &=& \!\sum_{\alpha=L,R}\sum_{k_\alpha} V_{k_\alpha s}^* f_{k_\alpha s} 
+ W_{12aa}^* d_{1a}+ W_{12ba}^* d_{1b}
\nnm \\
& &
+ W_{32aa}^* d_{3a}+ W_{32ba}^* d_{3b},
\nnm \\
(\omega -E_{k_\alpha}) f_{k_\alpha} &=& V_{k_\alpha,a}  d_{2a}+V_{k_\alpha,b} d_{2b}, \nnm 
\end{eqnarray}

From Eq.(\ref{d1aw}),
\begin{equation}
(\omega -E_{1a})\la d_{1a}^\dagger d_{1b} \ra
= W_{12aa}^* \la d_{2a}^\dagger d_{1b} \ra + W_{12ab}^* \la d_{2b}^\dagger d_{1b} \ra, 
\end{equation}
which results in the construction of the conventional time-ordered Green function, expressed as follows:
\begin{equation}
(\omega -E_{1a}) G_{d_{1b}d_{1a}}^t
= W_{12aa}^* G_{d_{1b}d_{2a}}^t + W_{12ab}^* G_{d_{1b}d_{2b}}^t.
\end{equation}
According to Jauho's procedure~\cite{Jauho}, no interaction is included in the lead, and we have
\begin{eqnarray}
G_{d_{1b}d_{1a}}^t
&=& \bigl[ W_{12aa}^* G_{d_{1b}d_{2a}}^t + W_{12ab}^* G_{d_{1b}d_{2b}}^t\bigr] \Sigma_{1a}, 
\end{eqnarray}
where $\Sigma_{1a}\equiv 1/(\omega-E_{1a})$.
For simplicity, $t$ is omitted from the shoulder of $G_{d_{1b}d_{1a}}^t$ in the following.
We then obtain the equations of the Green functions as follows:
\begin{eqnarray}
G_{d_{1b}d_{1a}}
&=& \bigl[W_{12aa}^* G_{d_{1b}d_{2a}} + W_{12ab}^* G_{d_{1b}d_{2b}} \bigr]\Sigma_{1a}, 
\nnm\\
G_{d_{1a}d_{1a}}
&=& \bigl[W_{12aa}^* G_{d_{1a}d_{2a}} + W_{12ab}^* G_{d_{1a}d_{2b}}+1\bigr]\Sigma_{1a}, 
\nnm\\
G_{d_{3a}d_{1a}}
&=& \bigl[W_{12aa}^* G_{d_{3a}d_{2a}} + W_{12ab}^* G_{d_{3a}d_{2b}} \bigr]\Sigma_{1a}, 
\nnm\\
G_{d_{3b}d_{1a}}
&=& \bigl[W_{12aa}^* G_{d_{3b}d_{2a}} + W_{12ab}^* G_{d_{3b}d_{2b}} \bigr]\Sigma_{1a},  
\nnm\\
G_{d_{2a}d_{1a}}
&=& \bigl[W_{12aa}^* G_{d_{2a}d_{2a}} + W_{12ab}^* G_{d_{2a}d_{2b}} \bigr]\Sigma_{1a},  
\label{G2a1a}\\
G_{d_{2b}d_{1a}}
&=& \bigl[W_{12aa}^* G_{d_{2b}d_{2a}} + W_{12ab}^* G_{d_{2b}d_{2b}}\bigr] \Sigma_{1a}.
\label{G2b1a}
\end{eqnarray}

Regarding the Green function of the electrodes, we obtain
\begin{equation}
(\omega -E_{k_\alpha a}) f_{k_\alpha}  V_{k_\alpha,a}  d_{2a}+V_{k_\alpha,b} d_{2b}.
\end{equation}
Therefore,
\begin{equation}
(\omega-E_{k_\alpha a}) G_{d_{2a}f_{k_\alpha}}
   =  V_{k_\alpha,a}^*G_{d_{2a}d_{2a}}
       +V_{k_\alpha,b}^*G_{d_{2a}d_{2b}}.
\end{equation}
This results in
\begin{eqnarray}
 G_{d_{2a}f_{k_\alpha}}&=&
  \bigl[V_{k_\alpha,a}^*G_{d_{2a}d_{2a}}
       +V_{k_\alpha,b}^*G_{d_{2a}d_{2b}}\bigr]g_{k_\alpha a},  
\label{f2-1} \\
 G_{d_{2b}f_{k_\alpha}}&=&
  \bigl[V_{k_\alpha,a}^*G_{d_{2b}d_{2a}}
       +V_{k_\alpha,b}^*G_{d_{2b}d_{2b}}\bigr]g_{k_\alpha a},  
\nnm \\
 G_{d_{1a}f_{k_\alpha}}&=&
  \bigl[V_{k_\alpha,a}^*G_{d_{1a}d_{2a}}
       +V_{k_\alpha,b}^*G_{d_{1a}d_{2b}}\bigr]g_{k_\alpha a},  
\nnm \\
 G_{d_{1b}f_{k_\alpha}}&=&
  \bigl[V_{k_\alpha,a}^*G_{d_{1b}d_{2a}}
       +V_{k_\alpha,b}^*G_{d_{1b}d_{2b}}\bigr]g_{k_\alpha a},  
\nnm \\
 G_{d_{3a}f_{k_\alpha}}&=&
  \bigl[V_{k_\alpha,a}^*G_{d_{3a}d_{2a}}
       +V_{k_\alpha,b}^*G_{d_{3a}d_{2b}}\bigr]g_{k_\alpha a},  
\nnm \\
 G_{d_{3b}f_{k_\alpha}}&=&
  \bigl[V_{k_\alpha,a}^*G_{d_{3b}d_{2a}}
       +V_{k_\alpha,b}^*G_{d_{3b}d_{2b}}\bigr]g_{k_\alpha a},  
\nnm \\
 G_{d_{1a}f_{k_\alpha}}&=&
  \bigl[V_{k_\alpha,a}^*G_{d_{1a}d_{2a}}
       +V_{k_\alpha,b}^*G_{d_{1a}d_{2b}}\bigr]g_{k_\alpha a},  
\nnm \\
 G_{f_{k_\beta}f_{k_\alpha}}&=&
  \bigl[V_{k_\alpha,a}^*G_{f_{k_\beta}d_{2a}}
       +V_{k_\alpha,b}^*G_{f_{k_\beta}d_{2b}}+\delta_{k_\alpha,k_\beta}\bigr]g_{k_\alpha a}.  
\nnm 
\end{eqnarray}

\subsection{Equations of the channel-QD Green functions}
Regarding the Green functions of the channel-QD, we obtain
\begin{eqnarray}
(\omega -E_{2a})d_{2a}
&=& \!\sum_{\alpha=L,R}\sum_{k_\alpha} V_{k_\alpha a}^*  f_{k_\alpha a} 
+ W_{12aa}^* d_{1a} +W_{12ba}^* d_{1b}
\nnm \\
& &
+ W_{32aa}^* d_{3a} +W_{32ba}^* d_{3b}.
\nnm 
\end{eqnarray}
Therefore,
\begin{eqnarray}
\lefteqn{
G_{d_{1a}d_{2a}}= \bigl[ \!\sum_{\alpha=L,R}\sum_{k_\alpha} V_{k_\alpha,a} G_{d_{1a}f_{k_\alpha}}
+ W_{12aa} G_{d_{1a}d_{1a}}
}
\nnm \\
& &+ W_{12ba}G_{d_{1a}d_{1b}}
+ W_{32aa} G_{d_{1a}d_{3a}}+ W_{32ba}G_{d_{1a}d_{3b}}\bigr]\Sigma_{2a},
\nnm \\
& & G_{d_{1b}d_{2a}}= \!\bigl[\sum_{\alpha=L,R}\sum_{k_\alpha} V_{k_\alpha,a} G_{d_{1b}f_{k_\alpha}}
+ W_{12aa} G_{d_{1b}d_{1a}}
\nnm \\
& &+W_{12ba}G_{d_{1b}d_{1b}}
+ W_{32aa} G_{d_{1b}d_{3a}}+ W_{32ba}G_{d_{1b}d_{3b}}\bigr]\Sigma_{2a},
\nnm \\
& & G_{d_{3a}d_{2a}}= \!\bigl[\sum_{\alpha=L,R}\sum_{k_\alpha} V_{k_\alpha,a} G_{d_{3a}f_{k_\alpha}}
+ W_{12aa} G_{d_{3a}d_{1a}}
\nnm\\
& &+ W_{12ba}G_{d_{3a}d_{1b}}
+ W_{32aa} G_{d_{3a}d_{3a}}+ W_{32ba}G_{d_{3a}d_{3b}}\bigr]\Sigma_{2a},
\nnm \\
& & G_{d_{3b}d_{2a}}= \!\bigl[\sum_{\alpha=L,R}\sum_{k_\alpha} V_{k_\alpha,a} G_{d_{3b}f_{k_\alpha}}
+ W_{12aa} G_{d_{3b}d_{1a}}
\nnm\\
& &+ W_{12ba}G_{d_{3b}d_{1b}}
+ W_{32aa} G_{d_{3b}d_{3a}}+ W_{32ba}G_{d_{3b}d_{3b}}\bigr]\Sigma_{2a},
\nnm \\
& & G_{d_{2a}d_{2a}}\!= \!\bigl[\sum_{\alpha=L,R}\sum_{k_\alpha}\! V_{k_\alpha,a} G_{d_{2a}f_{k_\alpha}}
\!+ W_{12aa} G_{d_{2a}d_{1a}}
\nnm \\
& &\!+ W_{12ba}G_{d_{2a}d_{1b}}
\!+ W_{32aa} G_{d_{2a}d_{3a}}\!+\! W_{32ba}G_{d_{2a}d_{3b}}\!+\!1 \bigr]\Sigma_{2a},
\nnm \\
\label{G2a2a} \\
& &G_{d_{2b}d_{2a}}= \!\bigl[\sum_{\alpha=L,R}\sum_{k_\alpha} V_{k_\alpha,a} G_{d_{2b}f_{k_\alpha}}\Sigma_{2a}
+ W_{12aa} G_{d_{2b}d_{1a}}
\nnm \\
& &+ W_{12ba}G_{d_{2b}d_{1b}}
+ W_{32aa} G_{d_{2b}d_{3a}}+ W_{32ba}G_{d_{2b}d_{3b}}\bigr]\Sigma_{2a},
\nnm \\
\label{G2b2a} \\
& & G_{f_{k_\beta}d_{2a}}= \!\bigl[\sum_{\alpha=L,R}\sum_{k_\alpha} V_{k_\alpha,a} G_{f_{k_\beta}f_{k_\alpha}}
+ W_{12aa} G_{f_{k_\beta}d_{1a}}
\nnm \\
& &+ W_{12ba}G_{f_{k_\beta}d_{1b}}
+ W_{32aa} G_{f_{k_\beta}d_{3a}}+ W_{32ba}G_{f_{k_\beta}d_{3b}}\bigr]\Sigma_{2a}.
\nnm 
\end{eqnarray}

Using Eqs.(\ref{G2a1a}), and (\ref{G2b1a}) with other Green functions,
Eq.(\ref{G2a2a}) is transformed to: 
\begin{eqnarray}
\lefteqn{
G_{d_{2a}d_{2a}}\!
= \bigl[ \!\!\sum_{\alpha=L,R}\sum_{k_\alpha} V_{k_\alpha,a} G_{d_{2a}f_{k_\alpha}}
\!+\! W_{12aa} G_{d_{2a}d_{1a}}
}
\nnm \\
\!&+&\! W_{12ba}G_{d_{2a}d_{1b}}
\! +\! W_{32aa} G_{d_{2a}d_{3a}}
\!+\! W_{32ba}G_{d_{2a}d_{3b}}\!+\!1 \bigr] \Sigma_{2a}
\nnm \\
&=& \sum_{\alpha=L,R}\sum_{k_\alpha} V_{k_\alpha,a} 
(V_{k_\alpha,a}^*G_{d_{2a}d_{2a}}
+V_{k_\alpha,b}^*G_{d_{2a}d_{2b}} 
)g_{k_\alpha a}\Sigma_{2a}
\nnm \\
&+&
W_{12aa}(W_{12aa}^* G_{d_{2a}d_{2a}}+ W_{12ab}^* G_{d_{2a}d_{2b}})\Sigma_{1a}\Sigma_{2a} 
\nnm \\
&+& 
W_{12ba}(W_{12bb}^* G_{d_{2a}d_{2b}}+W_{12ba}^* G_{d_{2a}d_{2a}})\Sigma_{1b}\Sigma_{2a} 
\nnm \\
&+&
W_{32aa}(W_{32aa}^* G_{d_{2a}d_{2a}}+W_{32ab}^* G_{d_{2a}d_{2b}})\Sigma_{3a}\Sigma_{2a} 
\nnm \\
&+& W_{32ba}(W_{32bb}^* G_{d_{2a}d_{2b}}+W_{32ba}^* G_{d_{2a}d_{2a}})\Sigma_{3b}\Sigma_{2a} 
\!+\!\Sigma_{2a}
\nnm \\
&=&  
(G_{d_{2a}d_{2a}}
+G_{d_{2a}d_{2b}}  
)\Sigma_{fa}\Sigma_{2a}
\nnm \\
&+&
 G_{d_{2a}d_{2a}} A_{vaa}
+G_{d_{2a}d_{2b}} A_{vab}
\!+\!\Sigma_{2a}.
\end{eqnarray}
Therefore, we have
\begin{eqnarray}
& &  
G_{d_{2a}d_{2a}}\!
[1- \Sigma_{fa}\Sigma_{2a}-A_{vaa}]
=G_{d_{2a}d_{2b}}
[\Sigma_{fa}\Sigma_{2a}
+A_{vab}]
\!+\!\Sigma_{2a}.
\nnm \\ 
\label{eq2da} 
\end{eqnarray}

Here, we define 
\begin{eqnarray}
A_{vaa}
&\equiv & 
 (|W_{12aa}|^2 \Sigma_{1a} 
 +|W_{12ba}|^2 \Sigma_{1b} 
 +|W_{32aa}|^2 \Sigma_{3a} 
\nnm\\
& &
 +|W_{32ba}|^2 \Sigma_{3b})\Sigma_{2a}, 
 \nnm\\
A_{vbb}
&\equiv&(|W_{12bb}|^2 \Sigma_{1b} 
 +|W_{12ab}|^2 \Sigma_{1a} 
 +|W_{32bb}|^2 \Sigma_{3b} 
\nnm\\
& &
 +|W_{32ab}|^2 \Sigma_{3a})\Sigma_{2b}, 
 \nnm\\
A_{vab}
&\equiv&
 (W_{12aa}W_{12ab}^* \Sigma_{1a} 
+ W_{12ba}W_{12bb}^* \Sigma_{1b}
+ W_{32aa}W_{32ab}^* \Sigma_{3a} 
\nnm\\
& &
+ W_{32ba}W_{32bb}^* \Sigma_{3b})\Sigma_{2a},
\nnm\\
A_{vba}&\equiv&
 (W_{12bb}W_{12ba}^* \Sigma_{1b} 
+ W_{12ab}W_{12aa}^* \Sigma_{1a}
+ W_{32bb}W_{32ba}^* \Sigma_{3b} 
\nnm\\
& &
+ W_{32ab}W_{32aa}^* \Sigma_{3a})\Sigma_{2b}.
\end{eqnarray}
$W_{12aa}$ ($W_{12ab}$) represents the tunneling between 
the energy level $E_{1a}$ in QD1 and energy level $E_{2a}$ ($E_{2b}$) in channel-QD.
$W_{12ba}$ ($W_{12bb}$) represents the tunneling between the energy level $E_{1b}$ in QD1 and energy level $E_{2a}$ ($E_{2b}$) in channel-QD.
Here, we assume 
\begin{eqnarray}
W_{12ab} &=&W_{12aa}, \ \ 
W_{12ba} =W_{12bb}, \nnm \\
W_{32ab} &=&W_{32aa}, \ \ 
W_{32ba} =W_{32bb}. \nnm 
\end{eqnarray}
Therefore, we have
\begin{eqnarray}
A_{vab}
&=&
 (|W_{12aa}|^2 \Sigma_{1a} 
+ |W_{12bb}|^2 \Sigma_{1b}
+ |W_{32aa}|^2 \Sigma_{3a} 
\nnm\\
& &
+ |W_{32bb}|^2 \Sigma_{3b})\Sigma_{2a}
=A_{vaa},
\nnm\\
A_{vba}&=&
 (|W_{12bb}|^2 \Sigma_{1b} 
+ |W_{12aa}|^2 \Sigma_{1a}
+ |W_{32bb}|^2 \Sigma_{3b} 
\nnm\\
& &
+ |W_{32aa}|^2 \Sigma_{3a})\Sigma_{2b}
=A_{vbb}.
\end{eqnarray}
Furthermore, we define 
\begin{eqnarray}
\Sigma_{f\xi}^< (\omega)&=& \sum_{ks}|V_{ks,\xi}|^2 g_{ks}^<(\omega) 
\nnm \\
&=& i [\Gamma^Lf_L(\omega)+\Gamma^R f_R(\omega)] =iF(\omega) \nnm \\
\Sigma_{f\xi}^r (\omega)&=& \sum_{ks}|V_{ks}|^2 g_{ks}^r(\omega), 
\nnm \\
&=& \sum_{ks}|V_{ks,\xi}|^2(P\frac{1}{\omega-E_{ks}}-i\pi \delta(\omega-E_{ks})) \nnm \\
&=& \sum_{ks}|V_{ks,\xi}|^2 P\frac{1}{\omega-E_{ks}}-i\gamma, \nnm
\end{eqnarray}
where $\xi=a,b$, and we take $\Sigma_{fa}=\Sigma_{fb}=\Sigma_{f}$.

Similarly, from Eq.(\ref{G2b2a}), we have
\begin{eqnarray}
\lefteqn{
G_{d_{2b}d_{2a}}
= \bigl[ \!\sum_{\alpha=L,R}\sum_{k_\alpha} V_{k_\alpha,a} G_{d_{2b}f_{k_\alpha}}
+ W_{12aa} G_{d_{2b}d_{1a}}
}\nnm \\
&+& W_{12ba}G_{d_{2b}d_{1b}}
+ W_{32aa} G_{d_{2b}d_{3a}}+ W_{32ba}G_{d_{2b}d_{3b}}\bigr] \Sigma_{2a}
\nnm \\
&=&  
[G_{d_{2b}d_{2a}} 
+G_{d_{2b}d_{2b}}] \Sigma_{fa}\Sigma_{2a}
+G_{d_{2b}d_{2a}}A_{vaa}+ G_{d_{2b}d_{2b}} A_{vab}.
\nnm
\end{eqnarray}
Therefore, we obtain
\begin{eqnarray}
& &  
G_{d_{2a}d_{2b}}[1-\Sigma_{fb}\Sigma_{2b}-A_{vaa}]
= G_{d_{2a}d_{2a}} [\Sigma_{fb}\Sigma_{2b}+ A_{vba}]. 
\nnm\\
\label{eq2db} 
\end{eqnarray}

\subsection{Solutions of $G_{d_{2a}d_{2a}}$, $G_{d_{2a}d_{2b}}$}
Here, we solve the Green functions of the channel-QDs. 
Eqs.(\ref{eq2da}) and (\ref{eq2db}) can be expressed as follows:
\begin{eqnarray}
& & G_{d_{2a}d_{2a}} b_{a} =G_{d_{2a}d_{2b}} a_{vab} +\Sigma_{2a},
\label{eq2da1}
\\
& & G_{d_{2a}d_{2b}}b_{b} = G_{d_{2a}d_{2a}} a_{vba},
\label{eq2db1} 
\end{eqnarray}
where
\begin{eqnarray}
b_a&\equiv& 1- \Sigma_{fa}\Sigma_{2a}-A_{vaa}, \nnm\\
b_b&\equiv& 1-\Sigma_{fb}\Sigma_{2b}-A_{vbb}, \nnm\\
a_{vab}&\equiv&\Sigma_{fa}\Sigma_{2a}+A_{vab},\nnm\\
a_{vba}&\equiv&\Sigma_{fb}\Sigma_{2b}+ A_{vba}.
\end{eqnarray}
Eqs.(\ref{eq2da1}) and (\ref{eq2db1}) can be solved directly as follows:
\begin{eqnarray}
G_{d_{2a}d_{2a}}&=& \frac{b_{b}\Sigma_{2a}}{b_{a}b_{b}- a_{vba}a_{vab}}, 
\nnm \\
G_{d_{2a}d_{2b}} &=& \frac{a_{vba}\Sigma_{2a}}{ b_{a}b_{b}- a_{vba}a_{vab}}. 
\end{eqnarray}
These forms have been utilized for restarted and advanced Green functions.
When the three Green functions, $A(E)$, $B(E)$, and $C(E)$ have the
relationship $A(E)=B(E)C(E)$, 
the lesser Green function $A^<(E)$ is given by~\cite{Jauho}
\begin{equation}
A^<(E) = B^r(E)C^<(E) + B^<(E)C^a(E).
\end{equation}
By applying this equation to Eqs.(\ref{eq2da1}) and (\ref{eq2db1}), we obtain
\begin{eqnarray}
& & G_{d_{2a}d_{2a}}^r b_a^< +G_{d_{2a}d_{2a}}^< b_a^a 
=G_{d_{2a}d_{2b}}^r a_{vab}^<+G_{d_{2a}d_{2b}}^< a_{vab}^a +\Sigma_{2a}^<,
\nnm \\
\label{eq2da2} \\
& & G_{d_{2a}d_{2b}}^r b_b^< +G_{d_{2a}d_{2b}}^< b_b^a = G_{d_{2a}d_{2a}}^r a_{vba}^<+G_{d_{2a}d_{2a}}^< a_{vba}^a. 
\label{eq2db2}
\end{eqnarray}

From Eq.(\ref{eq2db2}),
\begin{equation}
G_{d_{2a}d_{2b}}^< b_b^a = -G_{d_{2a}d_{2b}}^r b_b^< +G_{d_{2a}d_{2a}}^r a_{vba}^<
+G_{d_{2a}d_{2a}}^< a_{vba}^a, 
\end{equation}
resulting in
\begin{eqnarray}
G_{d_{2a}d_{2b}}^< 
&=& 
\Sigma_{2a}^r\frac{-a_{vba}^r b_b^<+b_b^r a_{vba}^<}{ b_a^r b_b^r- a_{vba}^r a_{vab}^r } \frac{1}{b_b^a} 
+G_{d_{2a}d_{2a}}^< a_{vba}^a\frac{1}{b_b^a}. 
\nnm\\
\label{q2a2b3}
\end{eqnarray}
By substituting this equation into Eq.(\ref{eq2da2}), we obtain 
\begin{eqnarray}
G_{d_{2a}d_{2a}}^< 
&=& \Sigma_{2a}^r 
 \frac{  -b_a^< |b_b^r|^2  +b_b^r a_{vba}^<a_{vab}^a
   +a_{vba}^r  a_{vab}^< b_b^a - b_b^< a_{vba}^r a_{vab}^a 
  }{ |b_a^r b_b^r- a_{vba}^r a_{vab}^r|^2} 
  \nnm\\
&+& \frac{\Sigma_{2a}^<  b_b^a}{(b_a^a b_b^a  -a_{vba}^a a_{vab}^a)  }.
\label{g2a2a3} 
\end{eqnarray}
From Eq.(\ref{q2a2b3}), we obtain
\begin{eqnarray}
G_{d_{2a}d_{2b}}^< 
&=& 
 \Sigma_{2a}^r
 \frac{
-a_{vba}^r b_b^< b_a^a
+ b_b^r b_a^a a_{vba}^< 
+(-b_a^< b_b^r  +a_{vba}^r  a_{vab}^<   ) a_{vba}^a
}{ |b_a^r b_b^r- a_{vba}^r a_{vab}^r|^2 } 
\nnm \\ 
 &+&\frac{\Sigma_{2a}^< a_{vba}^a}{(b_a^a b_b^a  -a_{vba}^a a_{vab}^a)  }.
 \label{g2a2b3} 
\end{eqnarray}

\begin{widetext}

\subsection{Derivation of the current formula}
By substituting Eq.(\ref{g2a2a3}) and Eq.(\ref{g2a2b3}) into Eq.(\ref{f2-1}), we obtain
\begin{eqnarray}
& & \sum_{k_L} V_{k_L,a} G_{d_{2a}f_{k_L}}^< 
=G_{d_{2a}d_{2a}}^r \Sigma_{faL}^<
+\Bigl(
 \Sigma_{2a}^r
 \frac{  -b_a^< |b_b^r|^2  +b_b^r a_{vba}^<a_{vab}^a
   +a_{vba}^r  a_{vab}^< b_b^a - b_b^< a_{vba}^r a_{vab}^a 
  }{ |b_a^r b_b^r- a_{vba}^r a_{vab}^r|^2} 
 +\frac{\Sigma_{2a}^<  b_b^a}{(b_a^a b_b^a  -a_{vba}^a a_{vab}^a)  }
\Bigr) \Sigma_{faL}^a
\nnm \\
& &    + G_{d_{2a}d_{2b}}^r \Sigma_{faL}^{<}
+\Bigl(
 \Sigma_{2a}^r
 \frac{
-a_{vba}^r b_b^< b_a^a
+ b_b^r b_a^a a_{vba}^<
+(-b_a^< b_b^r  +a_{vba}^r  a_{vab}^<   ) a_{vba}^a
}{ |b_a^r b_b^r- a_{vba}^r a_{vab}^r|^2 }  
 +\frac{\Sigma_{2a}^<a_{vba}^a}{(b_a^a b_b^a  -a_{vba}^a a_{vab}^a)  }
\Bigr)\Sigma_{faL}^{a}.  
\label{Gdf}
\end{eqnarray}
\end{widetext}
The first and third terms are expressed as follows:
\begin{equation}
 G_{d_{2a}d_{2a}}^r \Sigma_{fL}^<
+G_{d_{2a}d_{2b}}^r \Sigma_{fL}^{<}
=
 \Sigma_{2a}^r\frac{\Sigma_{fL}^<}  
{ b_{a}^rb_{b}^r- a_{vba}^r a_{vab}^r}.
\end{equation}
The numerators of the second and fourth terms in Eq.(\ref{Gdf}) are expressed as follows:
\begin{eqnarray}
& &
-b_a^< b_b^r (b_b^a + a_{vba}^a)  
+b_b^r a_{vba}^< (b_a^a+a_{vab}^a) 
\nnm \\
& & 
+a_{vba}^r  a_{vab}^< (b_b^a+a_{vba}^a) 
- b_b^< a_{vba}^r (b_a^a+a_{vab}^a)
\nnm \\
& &\rightarrow
(b_b^r +a_{vba}^r ) (a_{vab}^<  - b_b^<) 
\rightarrow
(a_{vab}^<  - b_b^<)
\nnm \\ 
& &=
 \Sigma_{fA}^r [x_{2a}^<+x_{2b}^<]
+\Sigma_{fA}^< [x_{2a}^a+x_{2b}^a].
\end{eqnarray}
Here, we have used
\begin{eqnarray}
b_a
&\rightarrow& 
1- \Sigma_{fva}\Sigma_{2a}, \nnm\\
b_b&\rightarrow& 
1- \Sigma_{fvb}\Sigma_{2b},\nnm\\
a_{vab}&\rightarrow& 
\Sigma_{fva}\Sigma_{2a},  \nnm\\
a_{vba}&\rightarrow& 
\Sigma_{fvb}\Sigma_{2b},
\end{eqnarray}
where
$\Sigma_{fva}\equiv \Sigma_{f} +{A}_{va}$
and
$\Sigma_{fvb}\equiv \Sigma_{f} +{A}_{vb}$.
Furthermore, we utilized
\begin{eqnarray}
b_b^a+a_{vba}^a&=& 1-\Sigma_{fb}^a \Sigma_{2b}^a -A_{vbb}^a +\Sigma_{fb}^a\Sigma_{2b}^a+ A_{vba}^a
= 1,
\nnm \\
b_a^a+a_{vab}^a &=& 1- \Sigma_{fa}^a\Sigma_{2a}^a -A_{vaa}^a +\Sigma_{fa}^a\Sigma_{2a}^a+A_{vab}^a
=1,
\nnm \\
 a_{vba}^<- b_a^<
&=&-(1- \Sigma_{fa}\Sigma_{2a}-A_{vaa} )^<
+(\Sigma_{fb}\Sigma_{2b}+ A_{vba})^<
\nnm \\
&=& 
 [\Sigma_{f}^r +{A}_{va}^r] \Sigma_{2a}^<
+[\Sigma_{f}^r +{A}_{vb}^r] \Sigma_{2b}^<
\nnm \\
& &
+[\Sigma_{f}^< +{A}_{va}^<] \Sigma_{2a}^a
+[\Sigma_{f}^< +{A}_{vb}^<] \Sigma_{2b}^a
\nnm \\
&=& 
 \Sigma_{fva}^r \Sigma_{2a}^<
+\Sigma_{fvb}^r \Sigma_{2b}^<
+\Sigma_{fva}^< \Sigma_{2a}^a
+\Sigma_{fvb}^< \Sigma_{2b}^a.
\nnm 
\end{eqnarray}

\begin{widetext}
Therefore, we have 
\begin{eqnarray}
& & \sum_{k_L} V_{k_L,a} G_{d_{2a}f_{k_L}}^< =
 \Sigma_{2a}^r\frac{\Sigma_{fL}^<}  
{ b_{a}^rb_{b}^r- a_{vba}^r a_{vab}^r}
+
\frac{ \Sigma_{fA}^r [x_{2a}^<+x_{2b}^<]
+\Sigma_{fA}^< [x_{2a}^a+x_{2b}^a]
}{ |b_a^r b_b^r- a_{vba}^r a_{vab}^r|^2 }\Sigma_{2a}^r\Sigma_{fL}^a
+
\frac{\Sigma_{2a}^<  }{(b_a^a b_b^a  -a_{vba}^a a_{vab}^a)  }
\Sigma_{fL}^a
\nnm \\
&=&
 \Sigma_{2a}^r
\Bigl( 
 \frac{
 \Bigl[\Sigma_{fL}^<(1-\Sigma_{fA}^a)
+\Sigma_{fA}^< \Sigma_{fL}^a
\Bigr][x_{2a}^a+x_{2b}^a]
+ \Sigma_{fA}^r \Sigma_{fL}^a[x_{2a}^<+x_{2b}^<]
}{ |b_a^r b_b^r- a_{vba}^r a_{vab}^r|^2 }
\Bigr)
+
\frac{\Sigma_{2a}^< }{(b_a^a b_b^a  -a_{vba}^a a_{vab}^a)  }
\Sigma_{fL}^a.
\label{ReIa}
\end{eqnarray}

\end{widetext}

The real part of the numerator is expressed as follows:
\begin{eqnarray}
& &
{\rm Re} \Sigma_{2a}^r
\Bigl(
\Sigma_{fL}^<(1-\Sigma_{fA}^a)
+\Sigma_{fA}^<\Sigma_{fL}^a 
\Bigr)[x_{2a}^a+x_{2b}^a]
\nnm \\
&\Rightarrow& 
 P\frac{1}{\omega-E_{2a}}
 \Bigl(
\Gamma_L f_L {\rm Im}\Sigma_{fA}^a
-F(\omega) {\rm Im}\Sigma_{fL}^a 
\Bigr) \nnm\\
&\times& [P\frac{1}{\omega-E_{2a}}+P\frac{1}{\omega-E_{2b}}]
\nnm \\
&=& 
 P\frac{1}{\omega-E_{2a}}[P\frac{1}{\omega-E_{2a}}+P\frac{1}{\omega-E_{2b}}]
[\frac{\Gamma_L\Gamma_R}{2} (f_L-f_R)].
\nnm
\end{eqnarray}
We assume $\Sigma_{fA}^<=\Sigma_f^<+A_v^<\Rightarrow \Sigma_f^<$. 
Terms, such as $\Sigma_{2a}^<$ can be neglected, because 
from the denominator $ |b_a^r b_b^r- a_{vba}^r a_{vab}^r|^2$, 
we have
$(\omega-E_{2a}) \Sigma_{2a}^<\propto (\omega-E_{2a}) \delta (\omega-E_{2a}) =0$.
Therefore, we obtain the current formula $I_{\rm D}$ expressed as follows:
\begin{eqnarray}
\lefteqn{
I_{\rm D}
= \frac{2e}{h} \int  d\omega 
{\rm Re}(\sum_{k_L} V_{k_L,a} G_{d_{2a}f_{k_L}}^<+\sum_{k_L} V_{k_L,b} G_{d_{2b}f_{k_L}}^<)
}
\nnm \\
&=&
\! \frac{e}{h} \!\int  d\omega 
\left[P\frac{1}{\omega-E_{2a}}+P\frac{1}{\omega-E_{2b}}\right]^2
 \frac{
\Gamma_L\Gamma_R(f_L-f_R)
}{ |b_a^r b_b^r- a_{vba}^r a_{vab}^r|^2 }.
\nnm \\
\end{eqnarray}
The denominator is expressed as follows:
\begin{eqnarray}
b_b b_a  -a_{vba} a_{vab}
&=&
(1- \Sigma_{fva}\Sigma_{2a}) (1- \Sigma_{fvb}\Sigma_{2b})
\nnm \\
& &
-\Sigma_{fva}\Sigma_{2a}  \Sigma_{fvb}\Sigma_{2b}
\nnm \\
&=& 1- \Sigma_{fva}\Sigma_{2a}- \Sigma_{fvb}\Sigma_{2b}.
\end{eqnarray}

\begin{widetext}
Therefore, the following equation is utilized
\begin{eqnarray}
& &
(1-\Sigma_{fa}^r\Sigma_{2a}^r-\Sigma_{fb}^r\Sigma_{2b}^r-A_{vaa}^r-A_{vbb}^r)
(\omega-E_{2a})(\omega-E_{2b})(\omega-E_{1a})(\omega-E_{1b})(\omega-E_{3a})(\omega-E_{3b})
\nnm \\
&\Rightarrow &
(\omega-E_{2a})(\omega-E_{2b})(\omega-E_{1a})(\omega-E_{1b})(\omega-E_{3a})(\omega-E_{3b})
  \nnm\\
& &
-[P\frac{|W_{12aa}|^2}{\omega-E_{1a}}
+P\frac{|W_{12bb}|^2 }{\omega-E_{1b}}
+P\frac{|W_{32aa}|^2}{\omega-E_{3a}}
+P\frac{|W_{32bb}|^2}{\omega-E_{3b}}]
 (\omega-E_{2b})(\omega-E_{1a})(\omega-E_{1b})(\omega-E_{3a})(\omega-E_{3b}) 
\nnm \\
& &
-[P\frac{|W_{12bb}|^2 }{\omega-E_{1b}}
+P\frac{|W_{12aa}|^2}{\omega-E_{1a}}
+P\frac{|W_{32bb}|^2}{\omega-E_{3b}}
+P\frac{|W_{32aa}|^2}{\omega-E_{3a}}]
 (\omega-E_{2a})(\omega-E_{1a})(\omega-E_{1b})(\omega-E_{3a})(\omega-E_{3b})
\nnm\\
& &
-i[ \gamma_a(\omega-E_{2b}) +  \gamma_b(\omega-E_{2a})](\omega-E_{1a})(\omega-E_{1b})(\omega-E_{3a})(\omega-E_{3b}).
\nnm
\end{eqnarray}

\end{widetext}


\begin{thebibliography}{99}
\bibitem{Google2023}
Google Quantum AI, 
Nature {\bf 614}, 676, (2023). 

\bibitem{IBM2023}
D. Castelvecchi,
Nature {\bf 624},238 (2023).

\bibitem{Fowler}
A.G. Fowler, M. Mariantoni, J.M. Martinis, and A.N. Cleland,
Phys. Rev. A {\bf 86}, 032324 (2012).

\bibitem{Loss}
D. Loss, and D.P. DiVincenzo, 
Phys.Rev. A {\bf 57}, 120 (1998).

\bibitem{Zwanenburg}
F. A. Zwanenburg, A. S. Dzurak, A. Morello, M. Y. Simmons, L. C. L. Hollenberg, G. Klimeck, S. Rogge, 
S. N. Coppersmith, and M. A. Eriksson
Rev. Mod. Phys. {\bf }85, 961 (2013).



\bibitem{TSMC2023}
S. Liao, L. Yang, T.K. Chiu, W.X. You, T.Y. Wu, K.F. Yang, W.Y. Woon, W.D. Ho, Z.C. Lin, H.Y. Hung,
J.C. Huang, S.T. Huang, M.C. Tsai, C.L. Yu, S.H. Chen, K.K. Hu, C.C. Shih, Y.T. Chen, C.Y. Liu, H.Y. Lin,
C.T. Chung, L. Su, C.Y. Chou, Y.T. Shen, C.M. Chang, Y.T. Lin, M.Y. Lin, W.C. Lin, B.H. Chen, C.S. Hou,
F. Lai, X. Chen, J. Wu, C.K. Lin, Y.K. Cheng, H.T. Lin, Y.C. Ku, S.S. Lin, L.C. Lu, S.M. Jang, and M. Cao
in 2023 International Electron Devices Meeting (IEDM).  29.6 (2023).

\bibitem{Intel2023}
C. J. Dorow, T. Schram, Q. Smets, K. P. O’Brien
, K. Maxey, C.-C. Lin, L. Panarella, B. Kaczer
, N. Arefin, A. Roy, R. Jordan, A. Oni, A. Penumatcha
, C. H. Naylor, M. Kavrik, D. Cott, B. Groven, V. Afanasiev, P. Morin
, I. Asselberghs, C. J. Lockhart de la Rosa, G. Sankar Kar, M. Metz, and  U. Avci
in 2023 International Electron Devices Meeting (IEDM).

\bibitem{IMEC2023}
I. Radu, B-Y. Nguyen, C-H. Chang, C. Roda Neve, G. Gaudin, G. Besnard, 
P. Batude, V. Loup, L. Brunet, A. Vandooren and
N. Horiguchi,
in 2023 International Electron Devices Meeting (IEDM).


\bibitem{Kim}
S. D. Kim, M. Guillorn, I. Lauer, P. Oldiges, T. Hook, and M.-H. Na,
Microelectron. Technol. Unified Conf. (S3S), 1, (2015). 

\bibitem{Ryckaert}
J. Ryckaert, M. H. Na, P. Weckx, D. Jang, P. Schuddinck, B. Chehab, S. Patli, S. Sarkar, O. Zografos, R. Baert,
and D. Verkest,
2019 IEEE International Electron Devices Meeting (IEDM), 29.4 (2019).




\bibitem{Michniewicz}
J. Michniewicz, and M. S. Kim 
Appl. Phys. Lett. {\bf 124}, 260502 (2024).

\bibitem{Elzerman}
J. M. Elzerman, R. Hanson, L. H. Willems van Beveren, B. Witkamp, L. M. K. Vandersypen,
and L. P. Kouwenhoven, 
Nature {\bf 430}, 431(2004).  

\bibitem{Koch}
M. Koch, J. G. Keizer, P. Pakkiam, D. Keith, M. G. House,
E. Peretz, and M. Y. Simmons,
Nature Nanotech 14, 137–140 (2019).

\bibitem{Keith}
D. Keith , M. G. House, M. B. Donnelly, T. F. Watson, B. Weber, and M. Y. Simmons,
Phys. Rev. X {\bf 9}, 041003 (2019).

\bibitem{tanaJAP}
T. Tanamoto and K. Ono, J. Appl. Phys. {\bf 134}, 214402 (2023).

\bibitem{Sankar}
D. Buterakos and S. Das Sarma,
PRX Quantum {\bf 2}, 040358 (2021).

\bibitem{Xuedong}
B. Tariq and X. Hu, 
npj Quantum Inf {\bf 8}, 53 (2022).

\bibitem{Goswami}
S. Goswami, K. A. Slinker, M. Friesen, L. M. McGuire, J. L. Truitt, C. Tahan, L. J. Klein, J. O. Chu, 
P. M. Mooney, D. W. van der Weide, Robert Joynt, S. N. Coppersmith, and M. A. Eriksson 
Nature Phys. {\bf 3}, 41 (2007). 


\bibitem{Dzurak}
C. H. Yang, A. Rossi, R. Ruskov, N. S. Lai, F. A. Mohiyaddin, S. Lee, C. Tahan, 
G. Klimeck, A. Morello, and A. S. Dzurak, 
Nat Commun {\bf 4}, 2069 (2013). 


\bibitem{Eriksson1}
Z. Shi, C. B. Simmons, D. R. Ward, J. R. Prance, X. Wu, T. S. Koh, J. K. Gamble, 
D. E. Savage, M. G. Lagally, M. Friesen, S. N. Coppersmith, and M. A. Eriksson,
Nat Commun {\bf 5}, 3020 (2014).

\bibitem{Hao}
X. Hao, R. Ruskov, M. Xiao, C. Tahan, and H. Jiang,
Nat Commun {\bf 5}, 3860 (2014). 

\bibitem{Eriksson2}
S. F. Neyens, R, H. Foote, B. Thorgrimsson, T. J. Knapp, T. McJunkin, L. M. K. Vandersypen, 
P. Amin, N. K. Thomas, J. S. Clarke, D. E. Savage, M. G. Lagally, M. Friesen, 
S. N. Coppersmith, and M. A. Eriksson,
Appl. Phys. Lett. {\bf 112}, 243107 (2018)

\bibitem{Ferdous} 
R. Ferdous, E. Kawakami, P. Scarlino, M. P. Nowak, D. R. Ward, D. E. Savage, M. G. Lagally, 
S. N. Coppersmith, M. Friesen, M. A. Eriksson, L. M. K. Vandersypen, and R. Rahman, 
npj Quantum Inf 4, 26 (2018). 
%


\bibitem{Zhang}
X. Zhang, R. Hu, H.-O. Li, F.-M. Jing, Y. Zhou, R.-L. Ma, M. Ni,
G. Luo, G. Cao, G.-L. Wang, X. Hu, H.-W. Jiang, G.-C. Guo, and G.-P. Guo,
Phys. Rev. Lett. {\bf 124}, 257701 (2020). 

\bibitem{Cai}
X. Cai, E. J. Connors, L. F. Edge, and J. M. Nichol, 
Nat. Phys. {\bf 19}, 386 (2023). 


\bibitem{Degli}
D. D. Esposti, L. E. A. Stehouwer, Ö. Gül, N. Samkharadze, 
C. D\'{e}prez, M. Meyer, I. N. Meijer, L. Tryputen, S. Karwal, 
M. Botifoll, J. Arbiol, S. V. Amitonov, L. M. K. Vandersypen, 
A. Sammak, M. Veldhorst, and G. Scappucci, 
npj Quantum Inf {\bf 10}, 32 (2024).

\bibitem{Smelyanskiy}
V. N. Smelyanskiy, A. G. Petukhov, and V. V. Osipov, 
Phys. Rev. B {\bf 72},081304 (2005). 

\bibitem{Culcer}
D. Culcer, A. L. Saraiva, B. Koiller, X. Hu, and S. Das Sarma, 
Phys. Rev. Lett. {\bf 108}, 126804 (2012)

\bibitem{Gamble}
J. K. Gamble, P. Harvey-Collard, N. T. Jacobson, A. D. Baczewski, E. Nielsen, L. Maurer,  
I. Montaño, M. Rudolph, M. S. Carroll, C. H. Yang, A. Rossi, A. S. Dzurak and R. P. Muller 
Appl. Phys. Lett. {\bf 109}, 253101 (2016).

\bibitem{Losert}
M. P. Losert, M. A. Eriksson, R. Joynt, R. Rahman, G. Scappucci, S. N. Coppersmith, and Mark Friesen
Phys. Rev. B {\bf 108}, 125405 (2023).

\bibitem{Bosco}
C. Adelsberger, S. Bosco, J. Klinovaja, and D. Loss,
Phys. Rev. Lett. {\bf 133}, 037001 (2024).


\bibitem{Jauho}
A.P. Jauho, N.S. Wingreen, and Y. Meir,
Phys. Rev. B {\bf 50}, 5528 (1994).

\bibitem{Kim2}
T. -S. Kim and S. Hershfield
Phys. Rev. B {\bf 63}, 245326 (2001).

\bibitem{tanaPRB}
T. Tanamoto and T. Aono, Phys. Rev. {\bf B} 106, 125401 (2022).

\bibitem{Takeda2}
K. Takeda, J. Kamioka, T. Otsuka, J. Yoneda, T. Nakajima, M.R. Delbecq,
S. Amaha, G. Allison,  T. Kodera,  S. Oda, and S. Tarucha,
Sci. Adv. {\bf 2}, e1600694 (2016).



\bibitem{Yoneda}
J. Yoneda, K. Takeda, A. Noiri, T. Nakajima, S. Li, J. Kamioka, T. Kodera, S. Tarucha, 
Nat. Communications, 
{\bf 11}, 1144 (2020).





\bibitem{Fei}
J. Fei, D. Zhou, Y. -P. Shim, S. Oh, X. Hu, and M. Friesen,
Phys. Rev. A {\bf 86}, 062328 (2012).

\bibitem{Burkard}
M. Russ, J. R. Petta, and G. Burkard, 
Phys. Rev. Lett. {\bf 121}, 177701 (2018).

\bibitem{Kandel}
Y. P. Kandel, H. Qiao, and  J. M. Nichol
Appl. Phys. Lett. {\bf 119}, 030501 (2021).

\bibitem{Noiri}
A. Noiri, K. Takeda, T. Nakajima, T. Kobayashi,
A. Sammak, G. Scappucci, and  S. Tarucha,
Nat Commun {\bf 13}, 5740 (2022). 



\bibitem{Medford}
J. Medford, J. Beil, J. M. Taylor, E. I. Rashba, H. Lu, A. C. Gossard, and C. M. Marcus
Phys. Rev. Lett. {\bf 111}, 050501 (2013).











\bibitem{Engel}
H.-A. Engel and D. Loss,
Phys. Rev. B {\bf 65}, 195321(2002).


\bibitem{Sanchez}
D. S\'{a}nchez, C. Gould, G. Schmidt, and L. W. Molenkamp,
IEEE Trans. Electron Devices {\bf 54}, 984 (2007).

\bibitem{Meir1}
Y. Meir, N.S. Wingreen, and P. A. Lee, 
Phys. Rev. Lett. {\bf 66}, 3048 (1991).


\bibitem{BSIM}
Y. S. Chauhan, D. D. Lu, V. Sriramkumar, S. Khandelwal, J. P. Duarte, N. Payvadosi, A. Niknejad, and  C. Hu,
{\it FinFET Modeling for IC Simulation and Design: Using the BSIM-CMG Standard}.
(Academic Press, London, 2015).

\bibitem{Schon}
Y. Makhlin, G. Sch\"{o}n, and A. Shnirman,
Rev. Mod. Phys. {\bf 73}, 357 (2001).

\bibitem{tanaAPL}
T. Tanamoto and K. Ono,
Appl. Phys. Lett. {\bf 119}, 174002 (2021).

\bibitem{Jena}
B. Jena, S. Dash, and G. P. Mishra, IEEE Trans. Electron Devices {\bf 65}, 2422
(2018).

\bibitem{Hill}
C. D. Hill, E. Peretz, S. J. Hile, M. G. House, M. Fuechsle, S. Rogge, M. Y. Simmons, 
and L. C.L. Hollenberg,
Sci. Adv. {\bf 1}, e1500707 (2015).

\bibitem{Veldhorst1}
M. Veldhorst, H. G. J. Eenink, C. H. Yang, and A. S. Dzurak,  
Nat. Commun. {\bf 8}, 1766 (2017).




\bibitem{Li}
L. Petit, D.P. Franke, J.P. Dehollain, J. Helsen, M. Steudtner, N.K. Thomas, 
Z.R. Yoscovits, K.J. Singh, S. Wehner, L.M.K. Vandersypen, J.S. Clarke, and M. Veldhorst,
Sci. Adv. {\bf 4}, eaar3960 (2018).







\bibitem{Veldhorst}
M. Veldhorst, C. H. Yang, J. C. C. Hwang, W. Huang, J. P. De
hollain, J. T. Muhonen, S. Simmons, A. Laucht, F. E. Hudson,
K. M. Itoh, A. Morello, and A. S. Dzurak, 
Nature {\bf  526}, 410 (2015).

 





\bibitem{tanaTCAD}
T. Tanamoto, 
2024 IEEE International 3D Systems Integration Conference (3DIC), 
Sendai, Japan, 2024.










\end{thebibliography}
\end{document}